\documentclass[12pt]{extarticle}%
\usepackage[a4paper, margin=1in]{geometry}
\usepackage{eurosym}
\usepackage{setspace}
\usepackage[utf8]{inputenc}
\usepackage{natbib}
\usepackage{amsmath,amsfonts,amssymb,amsthm,enumitem}
\usepackage{graphicx}
\usepackage{rotating}
\usepackage{subcaption}
\usepackage{multirow}
\usepackage{amsthm}
\usepackage{pgf}
\usepackage{setspace}
\usepackage{hyperref}
\usepackage{bbm}
\usepackage{comment}
\usepackage{booktabs}
\usepackage{tikz,calc}
\usepackage{accents}
\usepackage{xcolor}
\usepackage{soul}
\usepackage{pgfplots}
\usepackage{graphicx}
\usepackage{caption}
\usepackage{pdflscape}
\usepackage{float}
\usepackage{array}
\usepackage{multirow}
\usepackage{needspace}
\usepackage{placeins}
\usepackage{amsmath}
\usepackage{amsfonts}
\usepackage{xcolor}
\usepackage{amssymb}%
\providecommand{\U}[1]{\protect\rule{.1in}{.1in}}

\newtheorem{assumption}{Assumption}

\newtheorem{proposition}{Proposition}
\newtheorem{remark}{Remark}
\usepackage{authblk}
\DeclareUnicodeCharacter{2212}{-}

\begin{document}

\title{Network Effects in Corporate Emissions: Evidence from a Data-Dependent Spatial Panel Model}

\author[1]{Stylianos Asimakopoulos}
\author[2]{George Kapetanios \thanks{Corresponding author: George Kapetanios, King's Business School, King's College London, email: george.kapetanios@kcl.ac.uk}}
\author[1]{Vasilis Sarafidis}
\author[2]{Alexia Ventouri}

\affil[1]{Brunel University of London}
\affil[2]{King's College London}

\date{}
\maketitle

\begin{abstract}
We study spillover effects in corporate toxic emissions using a heterogeneous panel network of U.S.\ industrial facilities from 2000–2023. Rather than imposing a network structure a priori, we uncover an unobserved web of influence directly from the data using recent advances in high-dimensional network econometrics. Indirect effects transmitted through the estimated network account for about 28\% of the total impact of key firm balance-sheet characteristics. By contrast, distance-based networks generate no statistically discernible spillovers, while a priori firm- or industry-based networks substantially overstate within-group spillins relative to the data-driven network. These findings show that who is linked to whom, and with what strength, matters critically for assessing systemic environmental risk and for designing targeted regulation. Methodologically, the paper provides a flexible framework for quantifying facility-level emissions spillovers and their consequences in financial and policy settings.
\end{abstract}



\baselineskip 0.27in

\newpage

\section{Introduction}
Understanding the drivers of corporate emissions is a central concern for researchers and policymakers seeking to design effective climate mitigation strategies. While firm-level characteristics such as size, sector, technology adoption, and regulatory exposure play a critical role in shaping emissions outcomes \citep{cohen2018impact, dec2017}, firms do not operate in isolation. Their environmental decisions are influenced by competitors, supply chain partners, common investors, and regulators \citep{delmas2011us, mat2014, BJM2021, AVY2025, EJF}. These inter-firm interdependencies give rise to \textit{network or spillover effects}, whereby the emissions behaviour of one firm (or facility) affects that of others through complex and often unobserved channels.

Several mechanisms documented in prior theoretical work help to explain why emissions decisions may exhibit such interdependence. Firms compete on environmental reputation as well as price and quality, and may adjust emissions in response to peers’ abatement choices \citep{fow2012}. Cleaner technologies and production practices can diffuse along supply chains \citep{ace2012,car2014}. Institutional investors and lenders frequently coordinate engagement strategies across portfolio firms \citep{dyc2019,bol2021}. Environmental regulation, although typically defined at the sector or jurisdiction level, is often internalised at the group level and may trigger pre-emptive responses by neighbouring or competing firms \citep{gre2012,dec2017}. Taken together, these channels suggest that emissions decisions propagate through a rich, endogenous network of corporate interdependencies, not as isolated choices at the facility level.

Conceptually, if the relevant corporate linkages were directly observed, for example, who influences whom and with what intensity, one could recover spillover effects using standard spatial econometric techniques \citep{anselin1988spatial,lesage2009introduction,elhorst2014spatial,vega2015slx,CuiEtAl2023}. In practice, however, the underlying channels of influence rarely map cleanly into a quantifiable and well-defined notion of distance, and attempts to proxy them using e.g. simple geographic or industry-based metrics are necessarily ad hoc. Unfortunately, misspecifying the spatial weights matrix $\mathbf{W}$ can in turn lead to biased parameter estimates and misleading conclusions about the magnitude and distribution of spillover effects \citep{kelejian2010specification,chudik2011weak}.

The objectives of the present paper are twofold. First, to estimate a network of corporate emissions directly from the data, without imposing ex ante assumptions about which facilities are linked or the strength of their connections. Second, to investigate the forces that govern link formation and link strength in corporate emissions. In what follows, we refer to this as ``homophily'', by which we mean the tendency for units that are similar in terms of observable attributes or operating environment to be more likely to be connected and to exert stronger influence on one another’s emissions decisions.


In pursuing these objectives, we build on several recent contributions that recover sparse network structures from the data, for example \citet{LamSouza2016,LewbelEtAl2023,dePaulaEtAl2025}. However, these typically impose homogeneous slope coefficients and rule out ``interactive fixed effects''. Extending this line of work to relax these restrictions is particularly valuable in our setting for two main reasons. First, facilities and firms in our sample operate across a wide range of industries with distinct production technologies and emission profiles, and covariates are defined at the parent-firm level; thus, common slopes are likely to mask systematic heterogeneity both across industries and within firms. Second, two-way fixed effects cannot capture systematic heterogeneity in exposure to unobserved regulatory, energy-price, and reputational factors, as facilities in more energy-intensive or heavily regulated sectors, or with stronger ESG engagement by their parent firms, are expected to be more strongly affected by these latent drivers.
To this end, we adopt the novel ``Boosting One-Link-at-a-Time with Multiple Testing'' (BOLMT) procedure of \citet{JuodisEtAl2025}. The algorithm sequentially adds the most statistically significant links and uses a multiple-testing-based rule to determine when further link additions are no longer warranted. The procedure consistently recovers the network structure and, unlike much of the existing literature, accommodates heterogeneous slope coefficients across cross-sectional units, and interactive fixed effects.

Embedding the BOLMT procedure within our spatial panel framework yields a heterogeneous spatial panel model with an endogenously estimated network, $\widehat{\mathbf{W}}$. This set-up allows us to identify direct effects of firm characteristics on emissions, as well as indirect effects transmitted through the network, without imposing restrictive assumptions on latent heterogeneity and without specifying $\mathbf{W}$ ex ante. To the best of our knowledge, we are the first to analyse emissions data within a heterogeneous spatial panel data framework while constructing a data-dependent spatial weighting matrix via a stepwise multiple-testing procedure.

Our empirical setting is a panel of 399 industrial facilities, geographically dispersed across the United States and associated with 98 parent firms from 9 distinct industries, observed annually from 2000 to 2023. We combine facility-level emissions data from the U.S.\ EPA’s Toxics Release Inventory (TRI) with firm-level financial and governance information from Compustat and Execucomp. Emissions are modelled at the facility level, while balance-sheet and governance covariates are defined at the parent-firm level, allowing us to examine how firm-wide financial conditions and corporate policies propagate through a network of facilities.

Our main findings can be summarised in the following points:
\begin{enumerate}[label=(F\arabic*)]
    \item \textbf{Scale and efficiency as determinants of emissions.}
    Among our core balance-sheet covariates, total assets (which capture the scale of operations), exhibit a positive direct effect on emissions.
    By contrast, conditional on firm size, sales and capital expenditure (which proxy an efficiency margin in how existing capacity is operated and upgraded) display negative direct effects. This implies that firms with stronger revenue performance and greater investment activity tend to operate with lower emissions intensity.

    \item \textbf{Sparse and decentralized network.} The estimated network $\widehat{\mathbf{W}}$ is highly sparse with a density of $0.49\%$.\footnote{The total number of possible directed links is $N(N-1)$.
With $N = 399$, this yields $399 \cdot 398 = 158{,}802$ potential links.
Of these, $777$ links are estimated to be non-zero in $\widehat{\mathbf{W}}$,
implying a network density of $777/158{,}802 \approx 0.49\%$.} The maximum observed out-degree is 13 in a network of 399 nodes,
    and most nodes have only a small number of outgoing connections. This suggests the absence of hubs or dominant facilities. Instead, the network is sparse and decentralized, characterized by a relatively even distribution of influence across nodes. The lack of strong centralisation implies that emissions-related spillovers are not driven by a small set of key actors but are instead dispersed broadly across the network.

    \item \textbf{Evidence of substantial network spillovers.} We find strong and statistically significant network effects in corporate emissions behaviour. Indirect effects (those transmitted through the actions of interconnected facilities) account for roughly 28\% of the total marginal impact of key firm characteristics on emissions. In other words, a sizeable share of the influence of financial and operational variables is mediated by the network rather than arising solely from own-facility responses.
    This indicates that emissions behaviour cannot be fully understood at the level of individual firms, with important implications for targeted regulation, sustainable investing, and the pricing of climate-related risks in financial markets.

    \item \textbf{Geographic proximity is largely uninformative.} When the network is imposed ex ante using standard geographic weighting schemes based on physical distance, the implied indirect effects are small and statistically insignificant. This suggests that physical proximity is a poor proxy for the relevant channels of emissions interdependence among industrial facilities.

    \item \textbf{Limited contribution of spillins within same-firm, same-industry or same-state links.} We decompose the total indirect effect into contributions from links within the same firm, industry, or state and from links that span different firms, industries, and states, a decomposition that we refer to as ``spillins''. We find that same-industry links account for only a modest share of overall network propagation, with same-firm or same-state spillins even smaller. In other words, most of the indirect impact of firm characteristics on emissions is transmitted through links that extend beyond firm, industry, and state boundaries.

    \item \textbf{Evidence of size-based homophily.} Facilities are more likely to be linked when their parent firms are similar in size, as measured by total assets. In practice, this means that large firms tend to be connected to other large firms, and smaller firms to other small firms, even when they operate in different industries or locations. Spill-in decompositions by asset quintile indicate that facilities receive the majority of their indirect effects from links within the same size segment. For example, among facilities associated with firms in the top 20\% of the asset distribution, $83.1\%$ of the total indirect effect is transmitted through links from other large firms. 

    \item \textbf{A priori networks distort the pattern of spillovers.} We compare spillins obtained from the estimated network $\widehat{\mathbf{W}}$, with those implied by conventional firm-, industry-, or state-based networks. Imposing an industry-based network leads to much stronger within-industry propagation than is implied by the data-driven network. A regulator or investor relying on such a network would overstate the reach of industry-focused interventions through network channels and understate the importance of spillovers that operate across industries or along other dimensions such as firm size. More generally, network specifications that force links to lie entirely within firms, industries, or states substantially amplify the apparent importance of within-group spillins, whereas the empirically inferred network indicates that such group links account for only a modest share of total indirect effects.
\end{enumerate}

These results contribute to several strands of literature at the intersection of corporate behaviour and governance, environmental economics, and spatial and network econometrics. Methodologically, we demonstrate the feasibility and empirical value of estimating an endogenous network directly from the data, without imposing ex ante assumptions about which units are connected or the strength of those connections. Substantively, we find strong evidence of size-based homophily in the estimated network. This provides empirical support for theoretical arguments—previously largely untested in large-scale firm-level data—that link firm size to similarities in production technologies, capital intensity, and the scaling properties of environmental and abatement investments \citep[][]{jov1982,bar2000,por1995,bru2003}. The strong pattern of size-based homophily in spillins is consistent with the fact that benchmarking, disclosure practices, and financial and regulatory pressures are strongly stratified by firm size: large firms primarily compare environmental performance to other large, visible peers facing similar stakeholder and investor scrutiny, whereas smaller firms tend to emulate behaviour within their own size segment, under looser oversight and tighter financing constraints \citep{lea2014,fou2014,mat2014,bol2021,cla2008,mar2014,alb2019}.

The remainder of the paper is organised as follows. Section~\ref{sec:literature} reviews related work on emissions spillovers and spatial/network econometrics. Section~\ref{sec:Model} presents the model specification and motivation. Section~\ref{sec:estimation} develops the methodological framework and estimation strategy. Section~\ref{sec:findings} reports the main results, including network estimates, spillover decompositions, and robustness checks. Section~\ref{sec:network_formation} examines several potential mechanisms that govern link formation. A final section concludes.

\section{Literature Review}\label{sec:literature}

Research on corporate emissions has expanded rapidly over the past two decades, reflecting the growing importance of environmental performance in corporate strategy, investor decision-making, and regulatory policy. Within this literature, panel data models have become a core tool for analysing firm-level emissions over time. More recently, scholars have begun to recognise the importance of \textit{spillover effects}—both within and across industries—as a mechanism through which environmental behaviour propagates across firms. This section reviews four strands of work relevant to our analysis: (i) network effects and spillovers in corporate behaviour, (ii) corporate emissions and environmental performance, (iii) environmental signals in financial markets, and (iv) spatial and network econometrics.

\subsection{Network Effects and Corporate Spillovers}

A growing body of research documents that firms are influenced by the actions and characteristics of other firms.\footnote{In the panel data literature, this phenomenon is often referred to as cross-sectional dependence. See \citet{SarafidisWansbeek2012} and \citet{SarafidisWansbeek2021} for general treatments of this issue.} These network effects can arise through strategic competition, investor comparisons, benchmarking, and reputational concerns \citep{lea2014}. In the context of disclosure, \citet{Graham2005} show that firms condition financial communication and sustainability efforts on the behaviour of peers, while \citet{Jin2012} find that voluntary disclosure is shaped by competitive environments.
Similar patterns have been identified in innovation \citep{Bloom2013} and investment \citep{fou2014}, suggesting that corporate decisions are embedded in broader networks of interdependence. Our study extends this literature by focusing on environmental outcomes—specifically, toxic emissions—and by \emph{endogenising} the structure of firm connections, rather than relying on pre-specified network definitions.

\subsection{Corporate Emissions and Environmental Performance}

There is substantial evidence that firms’ environmental behaviour responds to regulatory, market, and social pressures. Early work such as \citet{konar2001does} and \citet{Greenstone2006} shows that pollution has tangible consequences for firm valuation and competitiveness. More recent studies emphasise the role of climate risk disclosure and environmental transparency in shaping firm outcomes \citep{Ilhan2023, bol2021}.
Most of this literature, however, treats firms as conditionally independent units and abstracts from inter-firm dynamics. Our contribution is to explicitly model the spillover structure among firms and to quantify the strength and reach of network effects in emissions behaviour.

\subsection{Environmental Signals in Financial Markets}

The growth of sustainable investing has heightened interest in how environmental performance feeds back into financial outcomes. \citet{alb2019}, \citet{Pastor2021}, and \citet{asimakopoulos2023role} find that firms with stronger environmental profiles enjoy higher valuations, greater investor demand, and better access to capital markets. \citet{Krueger2020} shows that institutional investors increasingly incorporate environmental metrics into capital allocation decisions.
These findings imply that environmental performance is both a real outcome and a financial signal. Network emissions behaviour may therefore affect a firm’s perceived sustainability profile, cost of capital, and stock performance. By endogenising the emissions network and identifying influential nodes, our framework speaks directly to how environmental spillovers shape the financial and regulatory landscape in which firms operate.

\subsection{Spatial and Network Econometrics}

Spatial econometric models provide a natural approach to account for cross-sectional dependence and spillover effects in panel data. Surveys and monographs such as \citet{anselin1988spatial} and \citet{lesage2009introduction} describe models that incorporate spatial lags of the dependent variable or the error term, governed by a spatial weighting matrix $\mathbf{W}$. These ideas have been extended to panel settings by \citet{elhorst2003specification, elhorst2014spatial,CuiEtAl2023}, among others, allowing spatial dependence to be modelled jointly with temporal dynamics.

A more recent literature seeks to \emph{estimate} $\mathbf{W}$ rather than impose it exogenously. \citet{qu2015estimating} develop an estimator for a spatial autoregressive model with an unknown $\mathbf{W}$, and \citet{bai2016matrix} propose matrix-completion methods to recover latent interaction structures. \citet{lam2019estimation}, \citet{zhang2020sparse} and \citet{dePaulaEtAl2025} use penalised regression and sparsity assumptions to estimate high-dimensional spatial structures from the data. Recently, \citet{JuodisEtAl2025} introduced a novel ``Boosting One-Link-at-a-Time with Multiple Testing'' procedure, which sequentially adds the most statistically significant links, combining stepwise selection with a multiple-testing-based stopping rule. Their approach consistently recovers the network structure. Moreover, unlike much of the existing literature, it allows for heterogeneous slope coefficients across cross-sectional units and interactive fixed effects.\footnote{Monte Carlo evidence in \citet{JuodisEtAl2025} suggests favourable finite-sample performance for the designs considered.}

\section{Model Specification and Motivation}
\label{sec:Model}

We model changes in toxic emissions using a panel of 399 distinct industrial facilities, observed annually from 2000 to 2023. These facilities are geographically dispersed across the United States and are associated with 98 parent firms.
In particular, we specify:
\begin{align}
y_{i,t} &= \psi_{i}
\sum_{j\neq i} w_{i,j} y_{j,t} + \boldsymbol{\beta}_{i}^{\prime} \boldsymbol{x}_{i,t} + u_{i,t}, \label{eq:dgp1} \\
u_{i,t} &= \boldsymbol{\lambda}_{i}^{\prime} \mathbf{g}_{t} + \varepsilon_{i,t}, \label{eq:u}
\end{align}
for $i = 1,\dots,N$ and $t = 1,\dots,T$, with $N = 399$ and $T = 24$. The 399 facilities span nine broad industry groups: Mining; Food Products; Wood, Furniture, and Paper; Chemicals and Petroleum; Metal Products; Machinery, Electronics, and Instruments; Transport Equipment; Utilities; and Public Administration.\footnote{Details on the underlying SIC-code classification and sample shares are reported in Table~\ref{tab:sic_9categories} in the Appendix.}

The dependent variable, $y_{i,t} \equiv \Delta \text{Total Emissions}_{i,t}$, is the first difference in industrial facility-level total emissions at year $t$.\footnote{Emission data are drawn from the U.S. Environmental Protection Agency’s Toxics Release Inventory (TRI), which records annual facility-level releases of listed toxic chemicals by establishments in covered industry sectors whose emissions exceed regulatory thresholds. Although TRI reports are self-reported, they are subject to extensive EPA verification and quality-assurance procedures. The TRI also provides historical facility identifiers and geo-codes (street address, latitude/longitude, and industry classification). For each facility–year, $\text{Total Emissions}_{i,t}$ is constructed as the sum of reported releases to air, water, and land (in million pounds).}
The spatial (network) lag term, $\sum_{j\neq i} w_{i,j} y_{j,t}$, is a weighted average of total emissions corresponding to facilities connected to $i$ (hereafter, ``links''), where $w_{i,j}$ denotes the $(i,j)$ element of the $N \times N$ adjacency (spatial-weights) matrix $\boldsymbol{W} = [w_{i,j}]$. $\boldsymbol{W}$ encodes who influences whom and with what intensity. The coefficient $\psi_{i}$ measures the strength of network spillovers in the emission decisions of facility $i$. A positive $\psi_{i}$ implies that facilities' abatement or emission choices are strategic complements, consistent with competitive benchmarking, technological diffusion, and coordinated investor or regulatory responses (see \citep{fow2012,bol2021,dec2017}), whereas a negative $\psi_{i}$ would be indicative of strategic substitution in emission behaviour.


The vector $\boldsymbol{x}_{i,t} = (x_{1,i,t}, x_{2,i,t}, \dots, x_{K,i,t})^{\prime}$ collects the following $K=7$ firm-level, time-varying balance-sheet characteristics:
\[
\begin{aligned}
x_{1,i,t} &\equiv \ln(\text{Assets}_{i,t-1}), &
x_{2,i,t} &\equiv \text{Leverage}_{i,t-1}, &
x_{3,i,t} &\equiv \ln(\text{Capital Expenditure}_{i,t-1}), \\
x_{4,i,t} &\equiv \text{Cash Ratio}_{i,t-1}, &
x_{5,i,t} &\equiv \ln(\text{Sales}_{i,t-1}), &
x_{6,i,t} &\equiv \text{Tobin’s Q}_{i,t-1}, \\
x_{7,i,t} &\equiv \text{Tangible Assets}_{i,t-1}.
\end{aligned}
\]
Detailed definitions and data sources for all variables used in our dataset are reported in Table \ref{tab:datadef}. These variables characterise firms' size, capital structure, liquidity, and growth opportunities, which are likely to shape both their exposure to environmental risks and their ability to undertake sustainability-related investments.

The choice of explanatory variables is informed by economic theory and prior evidence on the determinants of corporate environmental performance and pollution intensity. Firm size reflects scale effects and regulatory visibility: larger firms typically emit more due to their operational scope but are also subject to stronger scrutiny from regulators, investors, and the public, which can incentivize cleaner technologies \citep{konar2001does, mar2014}. Leverage proxies financial constraints that may limit investment in abatement or cleaner production: highly leveraged firms may prioritize short-term liquidity over long-term environmental compliance \citep{cao2019corporate}.
Capital expenditure proxies investment in productive assets and potentially environmental technologies. Higher capital spending may reduce emissions through modernization and energy efficiency, although existing evidence is mixed \citep{bushnell2013}. Cash holdings reflect financial flexibility and thus a firm’s capacity to absorb environmental costs and finance cleaner inputs; financially flexible firms are better placed to undertake emission-reducing initiatives \citep{alb2019}.
Sales control for market scale and customer visibility: firms with larger output may generate higher absolute emissions but can benefit from efficiency gains and face stronger stakeholder pressure to improve environmental performance \citep{delmas2008organizational}. Tobin’s Q summarises market valuation and growth opportunities; firms valued for future growth may invest more in sustainability to protect reputation and attract long-term investors \citep{Pastor2021}.
Finally, tangible assets proxy production intensity: firms with high tangibility ratios typically operate in capital-intensive sectors (e.g. manufacturing, energy), where emissions are inherently higher and technological spillovers across firms are more likely    .

Therefore, our specification links facility-level toxic emissions, measured using TRI data, to parent-firm-level financial covariates, constructed from consolidated balance sheets and reflecting firm-wide financial policies. This cross-level modelling strategy is standard in the literature linking corporate characteristics to facility-level environmental outcomes (see, for example, \citealp{BartramEtAl_2022}) and implies that facilities belonging to the same parent firm in a given year may differ in realized emissions due to local technology, regulatory conditions, or operational practices, while sharing a common firm-level financial environment.

In Section \ref{sec:robust}, we consider additional specifications that augment the baseline model in Eq. \eqref{eq:dgp1} by including governance-related controls:
\[
\begin{aligned}
x_{8,i,t}  & \equiv \text{Options Granted}_{i,t-1}; &
x_{9,i,t}  & \equiv \text{Female CEO}_{i,t-1}; \\
x_{10,i,t} & \equiv \text{Board Age Diversity}_{i,t-1}; &
x_{11,i,t} & \equiv \text{Executive Director}_{i,t-1}; \\
x_{12,i,t} & \equiv \text{CEO Remuneration}_{i,t-1}. &
\end{aligned}
\]
These variables are intended to capture internal decision-making structures that may shape environmental priorities \citep{haider2021corporate}.


From an econometric standpoint, it is crucial to allow for heterogeneous marginal effects in our model for at least two reasons. First, facilities and firms in our sample operate in a wide range of industries with distinct production technologies and pollution profiles; therefore, imposing common slopes is likely to mask systematic cross-industry heterogeneity in how financial and governance variables map into emission outcomes. Second, and perhaps more subtly, as the covariates are defined at the parent-firm level, the way in which a given balance-sheet or governance characteristic translates into emissions can differ across facilities within the same firm, reflecting variation in local technology, regulatory exposure, and operational practices.
Accordingly, the model in Eq.~\eqref{eq:dgp1} is specified as a heterogeneous panel data model. The heterogeneous spatial parameter $\psi_{i}$ and the vector of slope coefficients, $\boldsymbol{\beta}_{i}$, provide a more flexible and credible representation of the underlying emission–finance relationship and reducing the risk of bias from inappropriate slope homogeneity restrictions.

Finally, the error term $u_{i,t}$ is composite, consisting of $\varepsilon_{i,t}$, a purely idiosyncratic disturbance, and
$\boldsymbol{\lambda}_{i}^{\prime}\mathbf{g}_{t}$, a latent factor component (``interactive fixed effects'').
$\mathbf{g}_{t} = (g_{1,t}, \dots, g_{r_{y},t})^{\prime}$ denotes an $r_y \times 1$ vector of common shocks that affect all facilities, such as changes in environmental regulation, aggregate demand conditions, energy price shocks, or shifts in ESG-related investor and stakeholder pressure. The corresponding $r_y \times 1$ vector
$\boldsymbol{\lambda}_{i} = (\lambda_{1,i}, \dots, \lambda_{r_y,i})^{\prime}$ represents facility-specific factor loadings governing the strength of facility $i$’s exposure to these shocks. For instance, facilities in more emission-intensive, energy-intensive, or heavily regulated sectors are expected to exhibit larger loadings on regulatory and energy-price factors, whereas facilities with greater public
visibility or stronger ESG engagement by their parent firms may load more heavily on reputation- and investor-pressure factors.
This common factor structure encompasses the popular two-way error (TWE) components model as a special case, obtained by imposing $r_y = 2$, $g_{1,t} = 1$ for all $t$, and $\lambda_{2,i} = 1$ for all $i$.\footnote{Specifically, under these restrictions, $\boldsymbol{\lambda}{i}^{\prime}\mathbf{g}{t} = \lambda_{1,i} + g_{2,t}$, where $\lambda_{1,i}$ denotes a facility-specific fixed effect and $g_{2,t}$ captures common time effects that affect all facilities identically.}
As in the TWE specification, the factor component is intended to control for omitted variables and unobserved heterogeneity, but in a more general manner that allows for \textit{multiple} latent factors, \textit{heterogeneous} loadings, and rich forms of cross-sectional dependence in $u_{i,t}$.


Many of the aforementioned economy-wide forces that directly affect emissions, such as changes in environmental regulation, aggregate demand conditions, energy prices, and ESG-related investor and stakeholder pressure, also shape firms’ financial positions by influencing profitability, investment opportunities, and the cost of capital. At the same time, there may exist additional common shocks, such as monetary policy or broad credit conditions, that primarily affect financial variables but have no direct effect on emissions.
To accommodate these channels, we allow the firm-level covariates to be driven by common shocks and specify
\begin{equation}  \label{eq:dgp_x}
\boldsymbol{x}_{i,t} = \boldsymbol{\Gamma}_{i}^{\prime} \mathbf{f}_{t} + \mathbf{v}_{i,t},
\end{equation}
where $\mathbf{f}_{t} = (\mathbf{g}_{t}^{\prime}, \mathbf{f}_{x,t}^{\prime})^{\prime}$ is an $r \times 1$ vector of latent factors with $r = r_y + r_x$,  and $\mathbf{f}_{x,t}$ denoting the $r_x \times 1$ vector of additional shocks that operate primarily through the covariates. The matrix $\boldsymbol{\Gamma}_{i}$ is a $K \times r$ matrix of factor loadings, and $\mathbf{v}_{i,t}$ is a purely idiosyncratic $K \times 1$ component, assumed to be uncorrelated with $\varepsilon_{i,t}$ in \eqref{eq:u}.
Allowing $\mathbf{f}_{t}$ to contain both $\mathbf{g}_{t}$ and $\mathbf{f}_{x,t}$ therefore accommodates the possibility that some latent factors influence emissions both directly (through the term $\boldsymbol{\lambda}_{i}^{\prime}\mathbf{g}_{t}$ in \eqref{eq:u}) and indirectly via $\boldsymbol{x}_{i,t}$, while others operate only through the financial covariates.

\section{Estimation Approach} \label{sec:estimation}
Estimation of the model in Eq.~\eqref{eq:dgp1} faces two major complications.
First, the covariate vector $\boldsymbol{x}_{i,t}$ is endogenous because it contains
a latent common-factor component $\mathbf{f}_t$ that overlaps with $\mathbf{g}_{t}$ in $u_{i,t}$.
Second, the network matrix $\mathbf{W}$ is typically not observed in applications of this kind:
regulators and data providers do not record the strength of technological, regulatory, or
reputational spillovers between facilities, and simple ad hoc choices (such as geographic
distance or industry dummies) are unlikely to capture the relevant network structure.
Consequently, $\mathbf{W}$ must be recovered from the data.

To address both issues, we combine the Mean Group IV approach of \citet{ChenEtal2025} with the Boosting
One-Link-at-a-Time with Multiple Testing (BOLMT) procedure of \citet{JuodisEtAl2025}.

Stacking Eq. \eqref{eq:dgp1} and Eq. \eqref{eq:dgp_x}  over $t$ yields
\begin{equation}  \label{eq:dgp_vector}
\begin{split}
\mathbf{y}_{i}&=\psi_{i} \sum_{j\neq i}^{N} w_{i,j} \mathbf{y}_{j} +%
\mathbf{X}_{i}\boldsymbol{\beta}_i+\mathbf{u}_{i}; \\
\mathbf{u}_{i}&=\mathbf{G}\boldsymbol{\lambda}_{i}+\boldsymbol{\varepsilon}_{i}; \\
\mathbf{X}_{i}&=\mathbf{F}\boldsymbol{\Gamma}_{i}+\mathbf{V}_{i},
\end{split}%
\end{equation}
where  $\mathbf{y}_{i}=(y_{i,1},\ldots, y_{i,T})^{\prime }$, $\mathbf{y}_{j}=(y_{j,1},\ldots, y_{j,T})^{\prime}$, $\mathbf{X}_{i}=(\boldsymbol{x}_{i,1},\cdots,%
\boldsymbol{x}_{i,T})^{\prime }$, $\mathbf{G}=(\mathbf{g}_1,\ldots,\mathbf{g}%
_T)^{\prime }$ and $\boldsymbol{\varepsilon}_{i}=(\varepsilon_{i,1},\ldots,%
\varepsilon_{i,T})^{\prime }$, whereas $\mathbf{F}=(\boldsymbol{f}_1,\ldots,\boldsymbol{f}%
_T)^{\prime }$,  and $\mathbf{V}_{i}=(\mathbf{v}_{i,1},\ldots,%
\mathbf{v}_{i,T})^{\prime }$.

\subsection*{Defactoring}

To address endogeneity due to omitted variables and unobserved heterogeneity, we eliminate the latent factors by applying the
projection matrix $\mathbf{M}_{\mathbf{F}} = \mathbf{I}_T - \mathbf{F}\left(\mathbf{F}^{\prime }\mathbf{F}\right)^{-1}\mathbf{F}^{\prime}$ to Eq.~\eqref{eq:dgp_vector}:
\begin{equation}  \label{eq:dgp_defactored}
\mathbf{M}_{\mathbf{F}}\mathbf{y}_{i}
= \psi_{i} \mathbf{M}_{\mathbf{F}}\sum_{j\neq i}^{N} w_{i,j} \mathbf{y}_{j}
+ \mathbf{M}_{\mathbf{F}}\mathbf{X}_{i}\boldsymbol{\beta}_i
+ \mathbf{M}_{\mathbf{F}}\mathbf{u}_{i}.
\end{equation}
Since $\mathbf{M}_{\mathbf{F}}\mathbf{G} = \mathbf{0}$ and thereby $\mathbf{M}_{\mathbf{F}}\mathbf{u}_{i} = \mathbf{M}_{\mathbf{F}}\boldsymbol{\varepsilon}_{i}$, the defactored model in Eq.~\eqref{eq:dgp_defactored} can be written as
\begin{equation}  \label{eq:dgp_defactored_tilde}
\widetilde{\mathbf{y}}_{i}
= \psi_{i} \sum_{j\neq i}^{N} w_{i,j} \widetilde{\mathbf{y}}_{j}
+ \widetilde{\mathbf{X}}_{i}\boldsymbol{\beta}_i
+ \widetilde{\boldsymbol{\varepsilon}}_{i},
\end{equation}
where $\widetilde{\mathbf{y}}_{i} \equiv \mathbf{M}_{\mathbf{F}}\mathbf{y}_{i}$,
$\widetilde{\mathbf{X}}_{i} \equiv \mathbf{M}_{\mathbf{F}}\mathbf{X}_{i}$,
and similarly for the remaining variables.\footnote{In practice, $\mathbf{F}$ is unobserved, but its column space can be consistently estimated using principal components analysis, as in \citet{Bai2003}. In the implementation below, $\mathbf{M}_{\mathbf{F}}$ is therefore replaced by $\mathbf{M}_{\widehat{\mathbf{F}}}$.}

\subsection*{Recovering the network via BOLMT}

The BOLMT procedure of \citet{JuodisEtAl2025} is a stepwise selection method in which, for each unit $i$, network links are selected
sequentially and added one at a time using unit-specific IV regressions. Following \citet{chu2018}, a multiple-testing-based stopping rule determines when to stop adding links.

To see this, rewrite Eq. \eqref{eq:dgp_defactored_tilde}  as
\begin{equation}
\widetilde{\mathbf{y}}_{i}
= \sum_{s=1}^{k_{i}}\omega_{i,j_{s}}\widetilde{\mathbf{y}}_{j_{s}}
+ \boldsymbol{\beta}_{i}^{\prime}\widetilde{\mathbf{X}}_{i}
+ \widetilde{\boldsymbol{\varepsilon}}_{i},
\label{eq:intromodel}
\end{equation}
where setting $\omega_{i,j_{s}} = \psi_{i} w_{i,j_{s}}$ is due to the fact that $\psi_{i}$ and $w_{i,j_{s}}$ cannot be separately identified without additional normalisations. It is assumed that for every $i$ there exists a subset
$\mathcal{S}_{n,i} = \{ j_{1},\ldots,j_{k_{i}}\} \subset \mathcal{S}_{-i}
\equiv \{1,\ldots,i-1,i+1,\ldots,N\}$ such that $\omega_{i,j_{s}}\neq 0$ if and only if
$j_{s} \in \mathcal{S}_{n,i}$. The variables
$y_{j_{1},t},y_{j_{2},t},\ldots,y_{j_{k_{i}},t}$ are the $k_{i}$ unknown \emph{true} links
(or $n$-links) of $y_{i,t}$.
In addition to the $k_{i}$ true links, \citet{JuodisEtAl2025} distinguish two further subsets of
$\mathcal{S}_{-i}$. One, denoted by $\mathcal{S}_{d,i}$, consists of \emph{distant} units (or
$d$-units) that have zero weight $\omega_{i,j}$ and zero correlation with the $n$-links once the
effects of $\widetilde{\boldsymbol{X}}_{i}$ are filtered out. The remaining set, $\mathcal{S}_{p,i}$, contains
variables with zero weight $\omega_{i,j}$ but non-zero correlation with the true links conditional
on $\widetilde{\mathbf{X}}_{i}$. Elements of $\mathcal{S}_{p,i}$ are referred to as \emph{pseudo} or
\emph{proxy} links (or $p$-links). These correspond to connections that may be mistakenly interpreted as genuine links, and their presence complicates the analysis because they can obscure the identification of the true network structure.

The BOLMT procedure considers, for each $i$, the individual-specific IV regression of
$\widetilde{\mathbf{y}}_{i}$ on $\widetilde{\mathbf{X}}_{i}$, and $\widetilde{\mathbf{y}}_{j}$ one at a time, for
$j \neq i$, where the instruments for
$\widetilde{\mathbf{y}}_{j}$ are given by $\widetilde{\mathbf{X}}_{j}$. Let $t_{i,j}$ denote the $t$-ratio of the coefficient of $\widetilde{\mathbf{y}}_{j}$, for $i = 1,\ldots,N$ and $j \neq i$:
\begin{equation}
t_{i,j}
= \frac{T^{-1/2}\,\widehat{\widetilde{\mathbf{y}}}_{j}^{\prime}
\mathbf{M}_{\widetilde{\mathbf{X}}_{i}}\widetilde{\mathbf{y}}_{i}}
{\hat{\sigma}_{i}\sqrt{T^{-1}\,\widehat{\widetilde{\mathbf{y}}}_{j}^{\prime}
\mathbf{M}_{\widetilde{\mathbf{X}}_{i}}\widehat{\widetilde{\mathbf{y}}}_{j}}}\,,
\label{ti}
\end{equation}
where $\widehat{\widetilde{\mathbf{y}}}_{j}
= \mathbf{P}_{\widetilde{\mathbf{X}}_{j}}\widetilde{\mathbf{y}}_{j}$ is the
$T\times 1$ vector of fitted values of $\widetilde{\mathbf{y}}_{j}$,
$\mathbf{P}_{\widetilde{\mathbf{X}}_{j}}
= \widetilde{\mathbf{X}}_{j}\left( \widetilde{\mathbf{X}}_{j}^{\prime}
\widetilde{\mathbf{X}}_{j}\right)^{-1}\widetilde{\mathbf{X}}_{j}^{\prime}$,
$\mathbf{M}_{\widetilde{\mathbf{X}}_{i}} = \mathbf{I}_{T}-\mathbf{P}_{\widetilde{\mathbf{X}}_{i}}$,
and $\hat{\sigma}_{i}$ is the standard error of the IV regression.

The BOLMT stepwise approach exploits the fact that large values of $t_{i,j}$ are more likely
to be associated with $n$-units than with $p$-units. Once all $t_{i,j}$ have been computed, they are
ranked by absolute magnitude and the individual corresponding to the largest $t_{i,j}$ is selected
as a link, provided that this largest t-ratio exceeds a threshold. The threshold is given by
$c_{p}(N,\delta)$, a critical value function that tends to infinity as $N$ increases, which ensures
that no $d$-units are selected with probability approaching one. It is defined as
\begin{equation}
c_{p}(N,\delta) = \Phi^{-1}\!\left( 1 - \frac{p}{2f(N,\delta)}\right),
\label{cvf}
\end{equation}
where $\Phi^{-1}(\cdot)$ is the inverse standard normal distribution function,
$f(N,\delta) = cN^{\delta}$ for some positive constants $c$ and $\delta$, and
$p$ ($0 < p < 1$) is the nominal size of the individual tests chosen by the investigator.

After a link is selected, the corresponding $\widetilde{\mathbf{y}}_{j}$ is added to the set of
regressors for unit $i$, and the above steps are repeated (with appropriate instrumentation) until
no $t$-ratio exceeds the threshold $c_{p}(N,\delta)$. The algorithm stops when no further link is
selected for any $i$, and $\widehat{k}_{i}$ denotes the estimated number of links for each $i$.

A formal exposition of the algorithm is provided in the Appendix. The following proposition
summarises its selection properties. Let $\mathcal{A}_{0}$ denote the event that, for each $i$, we select all its true $n$-links and none of its $d$-units.

\begin{proposition}
Consider the DGP \eqref{eq:dgp1}, where each unit $i$ has $k_{i}$ link units, $k_{i}^{\ast}$ proxy units,
and $N - k_{i} - k_{i}^{\ast}$ distant units. Let $c_{p}(N,\delta)$ be defined as in \eqref{cvf},
where $0 < p < 1$ is a significance level and $f(N,\delta) = cN^{\delta}$ for some $c > 0$ and
$\delta > 0$. Let $N,T \rightarrow \infty$ such that $T = O(N^{\kappa_{1}})$ for some $\kappa_{1} > 0$,
and assume that $k_{i}^{\ast}$ and $k_{i}$ are finite for all $i$. Then, under Assumptions~\ref{ass1}–\ref{ass2} in the Appendix,
\begin{equation}
\lim_{N,T \rightarrow \infty} \Pr\left( \mathcal{A}_{0}\right) = 1.
\label{pA0}
\end{equation}
\end{proposition}

\begin{remark}
As stated in the proposition, the probability of $\mathcal{A}_{0}$ tends to one asymptotically. The critical value function $c_{p}(N,\delta)$ is constructed to diverge at a controlled rate so that, with probability approaching one, no $d$-unit is selected.
\end{remark}

\subsection*{Final IV and MGIV estimation}

In the final step, Eq.~\eqref{eq:intromodel} is re-estimated using the selected links, and
$\widehat{\mathbf{W}}$ is constructed by standardising the estimated coefficients
$\widehat{\omega}_{i,1},\ldots,\widehat{\omega}_{i,\widehat{k}_{i}}$ so that
$\sum_{s=1}^{\widehat{k}_{i}}\widehat{\omega}_{i,j_{s}} = 1$ for each $i$.

Given $\widehat{\mathbf{W}}$, we subsequently consider
\begin{equation}  \label{eq:dgp_defactored_tilde_C}
\widetilde{\mathbf{y}}_{i}
= \widetilde{\mathbf{C}}_{i} \boldsymbol{\theta}_{i}
+ \widetilde{\boldsymbol{\epsilon}}_{i},
\end{equation}
where $\widetilde{\mathbf{C}}_{i} = \left(\sum_{j\neq i}^{N} \widehat{w}_{i,j} \widetilde{\mathbf{y}}_{j},
\widetilde{\mathbf{X}}_{i}\right)$ and
$\boldsymbol{\theta}_{i} = \left(\psi_{i},\boldsymbol{\beta}_{i}^{\prime}\right)^{\prime}$. Following \citet{ChenEtal2025}, the parameter vector $\boldsymbol{\theta}_{i}$ is consistently estimated
using the IV estimator
\begin{equation}\label{IVi}
\widehat{\boldsymbol{\theta}}_{IV,i}
= \left(\mathbf{A}_i^{\prime}\mathbf{B}_i^{-1}\widehat{\mathbf{A}}_i\right)^{-1}
\mathbf{A}_i^{\prime}\widehat{\mathbf{B}}_i^{-1}\widehat{\mathbf{c}}_{i},
\end{equation}
where
\begin{equation}\label{notation1}
\mathbf{A}_i = T^{-1}\widetilde{\mathbf{Z}}_{i}^{\prime}\widetilde{\mathbf{C}}_{i},
\quad
\mathbf{B}_i = T^{-1} \widetilde{\mathbf{Z}}_{i}^{\prime}\widetilde{\mathbf{Z}}_{i},
\quad
\mathbf{c}_{i} = T^{-1}\widetilde{\mathbf{Z}}_{i}^{\prime}\widetilde{\mathbf{y}}_{i},
\end{equation}
and $\widetilde{\mathbf{Z}}_{i}=\left(\widetilde{\mathbf{X}}_{i}, \widetilde{\mathbf{X}}_{j} \right)$ collects all instruments.\footnote{In practice, all defactored variables (those with a tilde) are constructed using
$\mathbf{M}_{\widehat{\mathbf{F}}}$ in place of $\mathbf{M}_{\mathbf{F}}$.}

We then estimate the population average effect $\boldsymbol{\theta} \equiv E(\boldsymbol{\theta}_{i})$ using the following spatial Mean Group IV estimator
\begin{equation}\label{eq:IVMG}
\widehat{\boldsymbol{\theta}}_{MGIV}
= \frac{1}{N}\sum_{i=1}^N\widehat{\boldsymbol{\theta}}_{IV,i}.
\end{equation}

\begin{proposition}
Consider the DGP \eqref{eq:dgp1}, where each unit $i$ has $k_{i}$ link units, $k_{i}^{\ast}$ proxy units,
and $N - k_{i} - k_{i}^{\ast}$ distant units. Then, as $N,T\to \infty$ and under
Assumptions~\ref{ass1}–\ref{ass2} listed in the Appendix, the MGIV estimator
$\widehat{\boldsymbol{\theta}}_{MGIV}$ satisfies
\[
N^{-1/2}\left( \widehat{\boldsymbol{\theta}}_{MGIV} - \boldsymbol{\theta}\right) = O_p(1),
\]
and is asymptotically normally distributed.
\end{proposition}

\section{Empirical Findings}\label{sec:findings}

Our benchmark estimation method is the Mean Group IV (MGIV) estimator described in Section~\ref{sec:estimation}. The resulting estimated network $\widehat{\mathbf{W}}$ is very sparse: its density is only $0.49\%$, meaning that only $0.49\%$ of all possible links are non-zero.
To compare and contrast the results obtained under $\widehat{\mathbf{W}}$, we consider nine additional spatial MGIV implementations, which differ only in the choice of network matrix. Six of these specifications are based on physical distance and encode different assumptions on how spatial proximity governs interaction strength. The remaining three define facility-level links through common membership in the same industry, state, or firm.

Specifically, let $d_{i,j}$ denote the great-circle distance between facilities $i$ and $j$, computed using the Haversine formula from their geographic coordinates.
The first three distance-based networks, denoted $W_{10\%}$, $W_{5\%}$, and $W_{1\%}$, set $w_{i,j} = 1/d_{i,j}$ if $d_{i,j} \le x$ miles and $w_{i,j}=0$ otherwise. Here $x$ equals the 10th, 5th, and 1st percentile, respectively, of the empirical distribution of $d_{i,j}$ across all pairs $(i,j)$. This specification imposes a hard cutoff: only sufficiently close pairs receive strictly positive weights, inversely proportional to distance, thereby capturing highly localized spillovers. The resulting densities are 10\%, 5\%, and 1\%, respectively.

The network $W_{\text{Gaussian}}$ defines weights via a Gaussian decay function, $w_{i,j} = \exp\left(-\frac{d_{i,j}^2}{2\sigma^2}\right)$, where $\sigma$ controls the rate of decay.\footnote{This bandwidth parameter is set equal to the sample standard
deviation of $d_{i,j}$ divided by 3.} This specification avoids discrete cutoffs and assigns positive, though rapidly diminishing, weights to all facility pairs. 
The resulting density is 7.54\%.\footnote{Density is computed using links with $w_{i,j} > 0.01$.}

The $k$-nearest-neighbour networks, denoted $W_{5\text{nn}}$ and $W_{2\text{nn}}$, connect each facility to its $k \in {5,2}$ geographically closest facilities, with $w_{i,j} = 1$ if $j$ is among the $k$ nearest links of $i$ and $w_{i,j}=0$ otherwise. This construction yields a fixed degree of connectivity for each facility while allowing distances to vary, enforcing symmetry in the network. The corresponding densities are 1.26\% and 0.50\%, respectively.

Finally, the remaining three networks, denoted $W_{\text{industry}}$, $W_{\text{state}}$, and $W_{\text{firm}}$, link each facility $i$ to all other facilities in the same industry, state, and firm, respectively. The densities of these matrices are 12.33\%, 3.47\%, and 2.24\%.
Additional details on the characteristics of all these networks are reported in Table~\ref{tab:weights}.

To assess the role of spatial spillovers, we also benchmark our spatial specifications against a set of non-spatial estimators that do not account for such facility-level interactions. First, we consider two within (two-way fixed effects) specifications, \textit{TWFE-firm} and \textit{TWFE-facility}, which include common time effects and either firm- or facility-specific fixed effects, respectively. Second, we employ the two-stage IV (\textit{2SIV}) estimator of \citet{CuiEtal2022}, which generalises the TWFE framework by allowing for a latent common factor structure. Finally, we report results from the Mean Group IV estimator of \citet{NorkuteEtal2021}, which further extends \textit{2SIV} by allowing for slope parameter heterogeneity across facilities.

\subsection{Point Estimates and Standard Errors}

Tables~\ref{tab:main_results_coeff_A}--\ref{tab:main_results_coeff_B} report point estimates and standard errors for population-average effects. The first column of Table~\ref{tab:main_results_coeff_A} presents the spatial MGIV estimator based on the estimated network $\widehat{\boldsymbol{W}}$.

Several patterns stand out. Firm size (log assets) has a positive and statistically significant coefficient of 0.266 (p$<0.01$), indicating that larger firms emit more toxic pollutants, plausibly reflecting their greater operational scale and complexity. This result is consistent with \citet{konar2001does} and \citet{mar2014}, who document a positive association between firm size and pollution intensity.

Sales are negatively associated with emissions, with an estimated coefficient of $-0.287$ (p$<0.01$). This finding supports an “efficiency-with-scale” interpretation: more successful firms appear to operate with greater environmental efficiency, potentially due to stronger stakeholder scrutiny, tighter internal monitoring, or economies of scale in the adoption of cleaner technologies \citep{delmas2008organizational}.

Capital expenditures also display a negative coefficient ($-0.070$, p$<0.10$), suggesting that investment activity is modestly associated with improved environmental performance. This is consistent with the idea that firms renewing or expanding their capital stock may simultaneously upgrade to cleaner or more efficient production technologies. At the same time, the marginal significance level indicates that not all capital expenditure is environmentally oriented, so the emissions response likely depends on the composition of the investment.

By contrast, leverage, cash holdings, Tobin’s Q, and asset tangibility do not exhibit statistically significant effects. This suggests that, conditional on the other controls and the network structure, debt structure and market-based valuation do not exert a strong influence on facility-level emissions in this framework.

Finally, the coefficient on the spatial lag of emissions is positive and precisely estimated (0.285, p$<0.01$). This indicates substantial network interactions: facilities tend to emit more when their network links emit more, consistent with strategic complementarities, information diffusion, or shared regulatory and technological environments in shaping emissions behaviour. This interpretation accords with the diffusion and mimicry mechanisms emphasised by \citet{elhorst2003specification}, \citet{qu2015estimating}, and \citet{lam2019estimation}, who document that firms adjust emissions in response to the behaviour of interconnected firms and to reporting practices.

The remaining columns of Table~\ref{tab:main_results_coeff_A} report estimates obtained under alternative network matrices. For firm size, the coefficient remains positive and statistically significant across all specifications, although its magnitude is somewhat smaller under $W_{\text{firm}}$. The coefficient on sales is likewise negative and statistically significant for most network choices, with $W_{\text{firm}}$ being the only case in which the effect becomes imprecisely estimated. For capital expenditures, the coefficient is negative in all specifications but attains significance at the $10\%$ level only under $W_{5\text{nn}}$ and $W_{\text{state}}$, indicating that inference for this variable is sensitive to the choice of network.

As both the definition and scaling of the spatial lag, $\sum_{j\neq i} w_{i,j} y_{j,t}$, differ across networks (e.g.\ in terms of density and links), the magnitudes of the associated coefficients are not directly comparable across columns. We therefore focus on sign and statistical significance here, and defer to the implied indirect (spillover) effects (which are directly comparable across models) in the next subsection. The estimated spatial lag coefficient is positive in all remaining specifications, but its significance varies with the network structure. Among the six geographically defined or nearest-neighbour networks ($W_{10\%}$, $W_{5\%}$, $W_{1\%}$, $W_{\text{Gaussian}}$, $W_{5\text{nn}}$, $W_{2\text{nn}}$), three yield spatial coefficients that are not statistically different from zero, and, where significance is detected, it is only at the 10\% or 5\% levels; in none of these cases is the coefficient significant at the 1\% level. By contrast, for the firm-, industry-, and state-based networks, the estimated spatial coefficient is positive and significant at the 1\% level throughout.

At the bottom of Table~\ref{tab:main_results_coeff_A}, we also report the estimated number $r$ of unobserved common factors in the interactive fixed effects; in all cases, we obtain $r=2$, and this estimate is stable across all network specifications. These two latent factors represent pervasive sources of unobserved heterogeneity that affect emissions across facilities, and operate over and above facility fixed effects and common time effects already included in the specification. This underscores the importance of modelling interactive fixed effects: approaches restricted to additive (two-way) fixed effects are unlikely to adequately control for latent heterogeneity and cross-sectional dependence in emissions.

Turning to the non-spatial estimators in Table~\ref{tab:main_results_coeff_B}, the conventional TWFE-firm and TWFE-facility specifications deliver coefficients that are generally small and statistically insignificant. The \textit{2SIV} estimator recovers a few significant effects (notably for capex, cash, and sales), but these are modest in magnitude and sensitive to the exclusion of spatial spillovers. By contrast, the non-spatial \textit{MGIV} specification produces results that are broadly consistent in sign with the spatial MGIV model; most notably, a positive and significant coefficient on firm size and a negative and significant coefficient on sales. However, in the absence of an explicit network structure, \textit{MGIV} ignores feedback operating through inter-facility links, so the estimated coefficients reflect only direct effects and systematically understate total marginal effects.

\subsection{Direct and Indirect Effects}
We now formalize the notions of direct and indirect effects implied by our spatial specification. To this purpose, we rewrite \eqref{eq:dgp1} in vector form by stacking across units $i$ at each $t$, rather than across time $t$ for each $i$ as in Section~\ref{sec:estimation}:
\begin{equation} \label{eq:dgp_vector_t}
\mathbf{y}_{(t)}
= \boldsymbol{\Psi} \mathbf{W} \mathbf{y}_{(t)}
+ \sum_{\ell=1}^{K} \mathbf{B}_{\ell}\mathbf{x}_{\ell,(t)}
+ \mathbf{u}_{(t)},
\end{equation}
where $\mathbf{y}_{(t)}=\left(y_{1,t},\ldots,y_{N,t}\right)^{\prime}$,
$\mathbf{x}_{\ell,(t)}=\left(x_{\ell,1,t},\dots, x_{\ell,N,t}\right)^{\prime}$,
$\boldsymbol{\Psi} = \mathrm{diag}(\psi_1,\dots,\psi_N)$,
$\mathbf{B}_{\ell} = \mathrm{diag}(\beta_{\ell,1},\dots,\beta_{\ell,N})$,
and $\mathbf{u}_{(t)}=(u_{1,t},\ldots,u_{N,t})^{\prime }$.
Let $\mathbf{S}(\boldsymbol{\Psi}) \equiv \mathbf{I}_{N}-\boldsymbol{\Psi}\mathbf{W}$.
Assuming $\mathbf{S}(\boldsymbol{\Psi})$ is invertible, the reduced-form representation of the model is
\begin{equation} \label{eq:reduced_form}
\mathbf{y}_{(t)}
= \mathbf{S}(\boldsymbol{\Psi})^{-1}
\left( \sum_{\ell=1}^{K}\mathbf{B}_{\ell}\mathbf{x}_{\ell,(t)} +\mathbf{u}_{(t)} \right).
\end{equation}

For regressor $\ell$, the relevant impact matrix is therefore
\[
\mathbf{A}_{\ell}(\boldsymbol{\Psi})
\equiv \mathbf{S}(\boldsymbol{\Psi})^{-1}\mathbf{B}_{\ell},
\]
so that the marginal effects of $x_{\ell,j,t}$ on $\mathbf{y}_{(t)}$ are governed by the entries of $\mathbf{A}_{\ell}(\boldsymbol{\Psi})$, rather than by the slope coefficients $\beta_{\ell,i}$ alone. The diagonal elements of $\mathbf{A}_{\ell}(\boldsymbol{\Psi})$ capture own-facility (direct) effects, while the off-diagonal elements capture cross-facility (indirect or spillover) effects.

We summarise these element-wise effects using population-average impact measures. Let $\mathbf{1}_N$ denote an $N\times 1$ vector of ones. For regressor $\ell$, the average direct effect is defined as
\begin{equation}
\text{DE}_{\ell}(\boldsymbol{\Psi}) = \frac{1}{N}\,\mathrm{tr}\!\left(\mathbf{A}_{\ell}(\boldsymbol{\Psi})\right),
\end{equation}
i.e.\ the average of the diagonal elements of $\mathbf{A}_{\ell}(\boldsymbol{\Psi})$. The average total effect is given by
\[
\text{TE}_{\ell}(\boldsymbol{\Psi})
= \frac{1}{N}\,\mathbf{1}_N^{\prime}\mathbf{A}_{\ell}(\boldsymbol{\Psi})\mathbf{1}_N,
\]
which corresponds to the average row sum of $\mathbf{A}_{\ell}(\boldsymbol{\Psi})$. The average indirect (spillover) effect is then obtained as the difference
\[
\text{IE}_{\ell}(\boldsymbol{\Psi})
= \text{TE}_{\ell}(\boldsymbol{\Psi}) - \text{DE}_{\ell}(\boldsymbol{\Psi}).
\]
In what follows, we evaluate these impact measures based on the MGIV estimates of the population-average parameters.

Tables~\ref{tab:impact_decomposition_DE}–\ref{tab:impact_decomposition_TE} report the decomposition of marginal effects into direct, indirect, and total components for the spatial MGIV estimators. We focus initialy on the specification based on $\widehat{\mathbf{W}}$ (first column). For firm size, proxied by log assets, the direct effect is positive and statistically significant (0.267, $p<0.01$), and the corresponding total effect is 0.372 ($p<0.01$). Thus, direct effects account for roughly $72\%$ of the total effect, with the remaining $28\%$ (0.105) arising from spillovers transmitted through the network. Thus, larger firms not
only emit more themselves but also raise emissions among interconnected facilities.

In contrast, for sales the direct and indirect effects are both negative and statistically significant ($-0.288$, $p<0.01$, and $-0.114$, $p<0.01$, respectively), which yield a total effect of $-0.402$ ($p<0.01$). A one standard deviation increase in sales (1.615, see Table~\ref{tab:datadef}) is associated with a direct reduction in emissions of approximately $-0.465$ ($-0.288 \times 1.615$). When the indirect effect is taken into account, the implied reduction rises to $-0.649$ ($-0.402 \times 1.615$), i.e.\ about 28\% larger once spillovers are included. This suggests that firms with larger revenue bases tend to operate more cleanly and simultaneously exert a disciplining influence on the emissions behaviour of their network neighbours. Possible mechanisms include greater capacity to invest in cleaner technologies, stronger stakeholder scrutiny, or more stringent internal ESG standards.

Capital expenditures exhibit a smaller but still meaningful role, with a direct effect of $-0.071$ ($p<0.10$), an indirect effect of $-0.028$ ($p<0.10$), and a total effect of $-0.098$ ($p<0.10$). The combination of higher sales and capital expenditures may proxy for broader operational scale, allowing fixed-cost investments in abatement to be spread more efficiently. The indirect effects then imply that such facilities can affect environmental norms or practices among connected facilities, for example through market leadership, visibility, or shared supply-chain or ownership structures.

None of the remaining accounting variables, including Tobin’s Q, exhibits statistically significant spillover effects, which is consistent with the weak or insignificant direct effects for these covariates.

Comparing across network specifications, the pattern of indirect effects varies markedly with how links are defined. For the six geographically defined or nearest-neighbour matrices ($W_{10\%}$, $W_{5\%}$, $W_{1\%}$, $W_{\text{Gaussian}}$, $W_{5\text{nn}}$, $W_{2\text{nn}}$), none of the estimated indirect effects are statistically significant at conventional levels for any of the covariates, so the estimated impacts in these cases operate almost entirely through the direct channel. This pattern suggests that geographic proximity and simple nearest-neighbour rules are poor proxies for the effective connections along which emissions-related spillovers propagate.
By contrast, significant spillovers arise when the network is based on firm, industry, or state group membership. Under $W_{\text{firm}}$, the indirect effects for assets and sales are relatively modest and only weakly significant. The industry-based network $W_{\text{industry}}$ generates the largest spillovers: the indirect effect of assets reaches 0.248 (p$<0.05$) and that of sales $-0.219$ (p$<0.10$), accounting for approximately $44\%$ of the total impact. This implies much stronger network propagation than under geographically defined matrices and the estimated network $\widehat{\mathbf{W}}$. State-level links in $W_{\text{state}}$ also produce significant, though smaller, indirect effects for assets (0.050, p$<0.10$) and sales ($-0.047$, p$<0.10$). The indirect effect of capital expenditures is not statistically significant under any of the firm-, industry-, or state-based networks.

\subsection{Network Estimation}
\begin{figure}[htbp]
  \begin{center}
    \makebox[\textwidth][c]{%
      \includegraphics[width=1.3\textwidth]{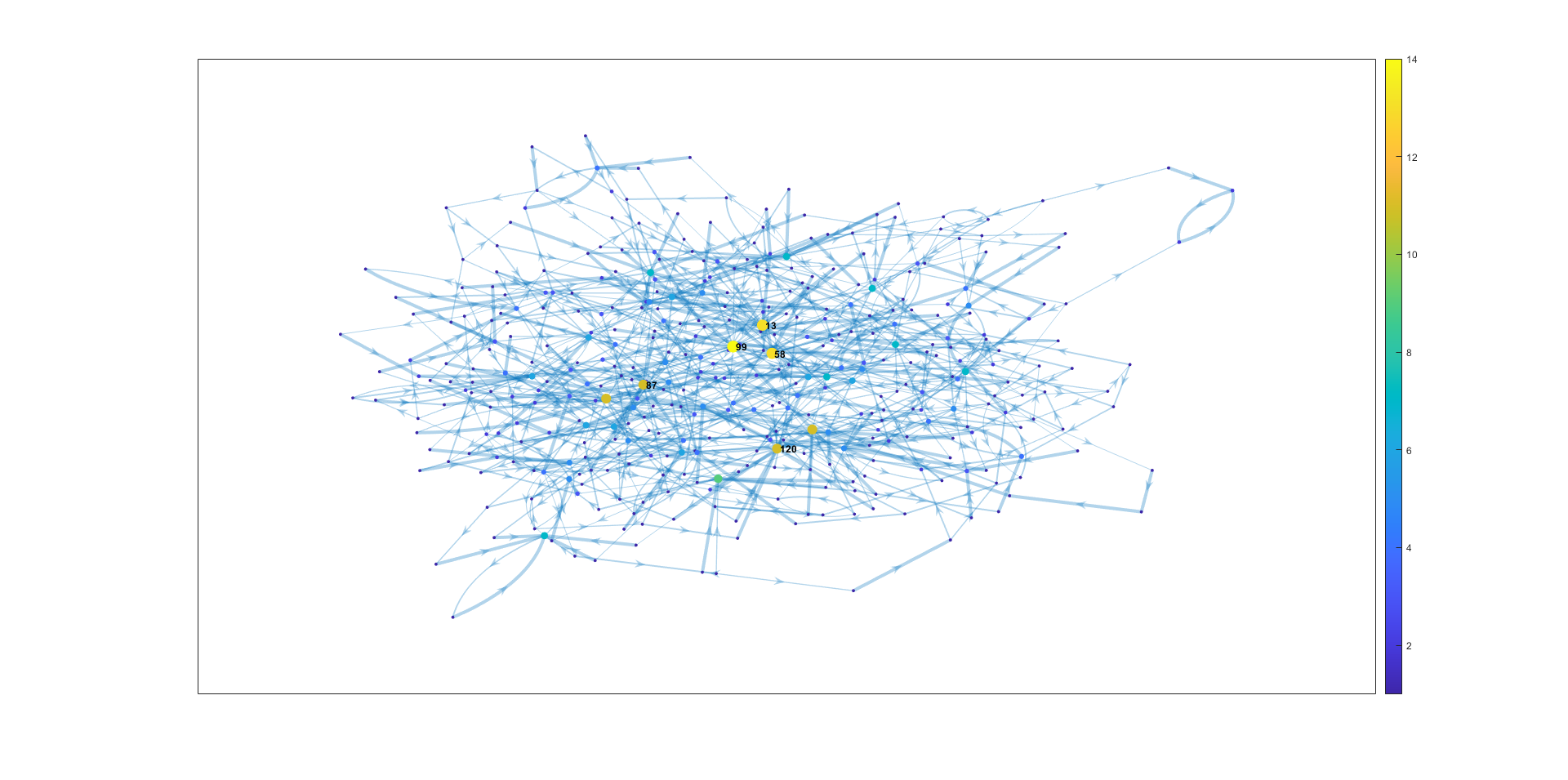}
    }
    \caption{\textbf{Directed Network Graph with Node Size and Color Scaled by Out-Degree}}
    \label{fig:W}
  \end{center}
\end{figure}

Figure~\ref{fig:W} visualizes the estimated network structure based on $\widehat{\boldsymbol{W}}$. Each node represents a facility, and edges capture estimated spillover links, i.e., the influence that one facility exerts on another in terms of emissions behavior. The network is plotted using a spring layout, which spatially clusters highly connected nodes toward the center and disperses more peripheral nodes outward.

Node size and color reflect out-degree centrality, the number of directed links a facility sends to others. While some variation is evident, the overall distribution of out-degrees is relatively narrow: the maximum observed out-degree is 13 in a network of 399 nodes, and most nodes have only a small number of outgoing connections. This suggests the absence of hubs or dominant facilities. Instead, the network is sparse and decentralized, characterized by a relatively even distribution of influence across nodes.

The lack of strong centralisation implies that emissions-related spillovers are not driven by a small set of key actors but are instead dispersed broadly across the network. This points to a form of environmental interdependence that is generated by many relatively small, local interactions rather than by a handful of dominant firms. The resulting structure highlights the importance of modelling network effects even in settings where no single unit exerts disproportionate influence, as aggregate spillovers may still emerge through indirect and cumulative linkages.

\section{Network Formation and Homophily Sources}\label{sec:network_formation}

While $\widehat{\mathbf{W}}$ captures the estimated adjacency matrix among facilities, it does not by itself reveal the mechanisms that govern link formation. That is, although it reflects the presence and intensity of connections, it remains silent on the economic, institutional, or geographic factors that might drive them. To explore these mechanisms, we proceed in three steps: we first examine homophily along firm, industry, and state dimensions, then investigate homophily in firm-specific balance-sheet characteristics, and finally analyse spillin effects under different network structures.

\subsection{Homophily along Firm, Industry, and State Dimensions}\label{subsection:homophily_category}

We examine whether facilities that belong to the same category (firm, industry or state) are more likely to form links than facilities in different categories, which would be consistent with category-based homophily in the network. For a generic grouping dimension $c \in \{\text{firm}, \text{industry}, \text{state}\}$, let $g_i^{c}$ denote the group label of facility $i$ along dimension $c$. In particular, $g_i^{\text{industry}} \in \{1,\dots,9\}$ denotes the industry classification (see Table~\ref{tab:sic_9categories}), $g_i^{\text{firm}} \in \{1,\dots,98\}$ the parent firm, and $g_i^{\text{state}} \in \{1,\dots,49\}$ the state.

For each $c \in \{\text{firm}, \text{industry}, \text{state}\}$, define
\begin{equation}
s^{c}_{i,j} =
\begin{cases}
1, & \text{if } g_i^{c} = g_j^{c} \text{ and } i \neq j, \\
0, & \text{otherwise},
\end{cases}
\end{equation}
so that $s^{\text{firm}}_{ij}$, $s^{\text{industry}}_{ij}$, and $s^{\text{state}}_{ij}$ indicate whether two facilities share the same parent firm, the same industry, or the same state, respectively.

The total number of same-category links observed in the estimated network $\widehat{\boldsymbol{W}}$ is then computed as:
\begin{equation}
L^{c}_{\text{same}} = \sum_{i \ne j} \widehat{w}_{i,j} \cdot s^{c}_{i,j},
\end{equation}
where $\widehat{w}_{i,j}$ denotes the $(i,j)$th entry of $\widehat{\boldsymbol{W}}$.

Thus, our first homophily index measures the proportion of total links that connect facilities facilities within the same category:
\begin{equation}
\widehat{h}^{c} = \frac{L^{c}_{\text{same}}}{L_{\text{total}}},
\end{equation}
where $L_{\text{total}} = \sum_{i \ne j} \widehat{w}_{i,j}$ is the total number of links in $\widehat{\mathbf{W}}$.

To assess whether the observed homophily is statistically significant, we evaluate $L^{c}_{\text{same}}$ (and, equivalently, $\widehat{h}^{c}$) using a permutation test, based on \citet{HannemanRiddle2005} and \citet{LaFondNeville2010}. Under the null hypothesis of no category-based clustering along dimension $c$, group labels are randomly assigned to facilities, independently of the network structure. The procedure is implemented separately for $c \in \{\text{firm}, \text{industry}, \text{state}\}$ as follows:

\begin{itemize}
    \item For a given $c$, we randomly permute the group labels $\{g_i^{c}\}_{i=1}^N$, generating a permuted label vector $\{g_{i}^{c,(b)}\}_{i=1}^N$ for each permutation $b = 1, \dots, B$.
    \item For each permutation, we construct the corresponding same-category indicators
    \[
    s^{c,(b)}_{ij} =
    \begin{cases}
    1, & \text{if } g_i^{c,(b)} = g_j^{c,(b)} \text{ and } i \neq j, \\
    0, & \text{otherwise}.
    \end{cases}
    \]
    \item We compute the total weight of same-category links for each permutation:
    \begin{equation}
    L_{\text{same}}^{c,(b)} = \sum_{i \ne j} \widehat{w}_{i,j} \, s^{c,(b)}_{ij}.
    \end{equation}
    \item The one-sided $p$-value for dimension $c$ is then estimated as
    \begin{equation}
    p^{c} = \frac{1}{B} \sum_{b=1}^{B} \mathbf{1} \left( L_{\text{same}}^{c,(b)} \geq L_{\text{same}}^{c} \right).
    \end{equation}
\end{itemize}
We set $B = 10{,}000$. A small $p^{c}$ indicates that the observed total weight of same-category links along dimension $c$ is significantly higher than that implied by random assignment, providing evidence of homophily by firm, industry, or state, respectively.

Under the null hypothesis of no category-based clustering in link formation along dimension $c$, the expected proportion of links, $\mathbb{E}[\hat{h}^{c}_{\text{null}}]$, can be obtained as:
\begin{equation}
\mathbb{E}[\hat{h}^{c}_{\text{null}}] = \frac{1}{B} \sum_{b=1}^{B} \frac{L^{c,(b)}_{\text{same}}}{L_{\text{total}}}.
\end{equation}
Thus, we also define a second index, the within-category homophily excess in percentage points (p.p.),
\begin{equation}
\widehat{h}^{c} - \mathbb{E}[\widehat{h}^{c}_{\text{null}}],
\end{equation}
which measures by how many percentage points the observed proportion of same-category links exceeds its expected value under random assignment.

The following output summarizes and reports results with respect to all these measures:
\begin{center}
\begin{tabular}{lccccc}
\hline
Category & $L_{\text{same}}^{c}$ & $\widehat{h}^{c}$ & $\mathbb{E}\!\left[\widehat{h}^{c}_{\text{null}}\right]$ & $\widehat{h}^{c} - \mathbb{E}\!\left[\widehat{h}^{c}_{\text{null}}\right]$ & $p^{c}$ \\
\hline
Firm     & 26  & 3.3\%  & 2.1\%  & 0.012 & 0.036 \\
Industry & 115 & 14.7\% & 12.3\% & 0.024 & 0.029 \\
State    & 31  & 4.0\%  & 3.5\%  & 0.005 & 0.258 \\
\hline
\end{tabular}
\end{center}

For firm-level homophily, we observe 26 within-firm links (out of 777 in total), corresponding to $\widehat{h}^{\text{firm}} = 3.3\%$ of all links, compared with an expected share of 2.1\% under random assignment. The implied excess of 1.2 percentage points is statistically significant ($p^{\text{firm}} = 0.036$), indicating that facilities belonging to the same parent firm are connected more often than would be expected by chance. Industry-level homophily is even stronger: 115 within-industry links yield $\widehat{h}^{\text{industry}} = 14.7\%$ versus a null benchmark of 12.3\%, an excess of 2.4 percentage points that is again statistically significant ($p^{\text{industry}} = 0.029$). By contrast, state-level homophily is modest: 31 within-state links correspond to $\widehat{h}^{\text{state}} = 4.0\%$ compared with 3.5\% under the null, an excess of only 0.5 percentage points that is not statistically significant ($p^{\text{state}} = 0.258$).Overall, these results indicate that homophily is most pronounced along industry groupings, followed by firm-level clustering, with comparatively weak evidence of geographic (state-level) clustering in the estimated network.

\subsection{Homophily in Financial Characteristics}\label{subsection:homophily_financial_characteristics}

We now examine potential homophily in firm-specific financial characteristics. Since these variables are not categorical, we adopt a different strategy. Let $\overline{\mathbf{x}}_{\ell} = (\overline{x}_{\ell,1}, \dots, \overline{x}_{\ell,N})^{\prime}$ denote the $N \times 1$ vector of time-averaged values of covariate $x_{\ell,i,t}$ across $t = 1,\dots,T$, where
\[
\overline{x}_{\ell,i} = \frac{1}{T} \sum_{t=1}^{T} x_{\ell,i,t}
\]
for each facility $i$ and variable $\ell = 1,\dots,K$. For each covariate $\ell$, we construct the $N \times N$ matrix of pairwise distances
\[
\mathbf{D}_{\ell}^{x} = \big[ d_{\ell,i,j}^{x} \big]_{i,j=1}^{N};
\quad d_{\ell,i,j}^{x} = \big|\overline{x}_{\ell,i} - \overline{x}_{\ell,j}\big|.
\]

We then vectorise the off-diagonal elements of both the estimated network and the distance matrix. Specifically, let
\[
\widehat{\mathbf{w}} = \operatorname{vec}_{\text{off}}(\widehat{\mathbf{W}});
\qquad
\mathbf{d}_{\ell}^{x} = \operatorname{vec}_{\text{off}}(\mathbf{D}_{\ell}^{x}),
\]
where $\operatorname{vec}_{\text{off}}(\cdot)$ stacks all $i \neq j$ entries of a matrix into a column vector. Indexing these off-diagonal pairs by $m = 1,\dots,M$ (with $M = N(N-1)$), let $\widehat{w}_{m}$ denote the $m$th element of $\widehat{\mathbf{w}}$ and $d^{x}_{\ell,m}$ the corresponding element of $\mathbf{d}_{\ell}^{x}$. We model link formation via the logistic regression
\begin{equation}
\label{eq:logit_homophily}
\log\left(\frac{\Pr(\widehat{w}_{m}=1 \mid \mathbf{d}^{x}_{m})}{1 - \Pr(\widehat{w}_{m}=1 \mid \mathbf{d}^{x}_{m})}\right)
= \alpha + \sum_{\ell=1}^{K} \delta_{\ell}\, d^{x}_{\ell,m},
\end{equation}
where $\mathbf{d}^{x}_{m} = (d^{x}_{1,m},\dots,d^{x}_{K,m})^{\prime}$ collects the pairwise distances for all covariates and
$\boldsymbol{\delta} = (\delta_{1},\dots,\delta_{K})^{\prime}$ is the corresponding coefficient vector.
The parameters $(\alpha,\boldsymbol{\delta})$ are estimated using a bias-corrected logistic estimator.
This approach allows us to assess whether link formation is systematically related to pairwise distances in financial characteristics, without imposing a rigid parametric specification for the decay pattern (such as an exponential kernel). Bias correction is important in our setting because the estimated network is sparse (the density of $\widehat{\mathbf{W}}$ is around 0.5\%), so conventional logistic estimates would otherwise be subject to substantial small-sample bias.

Among the candidate regressors, only one variable is found to be statistically significant at the 1\% level: the distance in log-transformed assets. The estimated coefficient on this regressor is $\widehat{\delta}_{\text{assets}} = -0.29$, with an associated odds ratio of $\exp(-0.29) \approx 0.748$. This implies that for every one-unit increase in the absolute difference in log assets between two facilities, the odds of a link between them decrease by approximately 25\%. In other words, two facilities that differ more in size (as measured by assets) are substantially less likely to be connected in the estimated network.\footnote{To further assess whether facilities with similar assets are more likely to be linked, we have implemented a nonparametric Wilcoxon rank-sum test comparing log-asset distances for linked and unlinked pairs. The test yields $z = -3.14$ with a $p$-value of $0.002$, indicating that distances in log assets are significantly smaller among linked facilities. This provides strong statistical evidence that linked facilities are more similar in size (as measured by assets), reinforcing the interpretation that link formation is partly driven by homophily in financial scale.}

\subsection{Spillins}
We next examine how indirect effects decompose across different groups of links. Following the spatial econometrics literature, we use the term spillin to denote the contribution of a particular subset of links (e.g.\ same-firm, same-industry, or same-size connections) to the overall indirect effect of a given covariate. For each grouping, we distinguish three types of spillins: within-group spillins (links connecting facilities in the same group), between-group spillins (links connecting facilities in different groups), and an all-link spillin, which matches the total indirect effect reported in Table~\ref{tab:impact_decomposition_IE}.

Table~\ref{tab:spillins} reports spillins for the spatial MGIV estimator based on the inferred network $\widehat{\mathbf{W}}$, decomposed by firm, industry, and state. Two features stand out. First, within-group spillins account for only a minor fraction of the total, especially when grouping by firm or state. Specifically, only 2.2\% of the total spillin arises from same-firm links and 3.9\% from same-state links. Even along the industry dimension, within-industry links generate just 13.1\% of the total spillin for assets and sales. Thus, consistent with the findings in Section \ref{subsection:homophily_category}, the bulk of the spillins implied by the estimated network are transmitted through inter-facility links that extend beyond firm, industry, and state boundaries.

Table~\ref{tab:spillins_quintile} partitions facilities into quintiles of the cross-sectional distribution of time-averaged assets and decomposes spillins into within- and between-quintile components. For each quintile $Q_j$, the ``Within'' entry captures the indirect effect arising from links to facilities in the same size quintile, whereas ``Between'' captures links to all other quintiles. Here the pattern is markedly different: within-quintile links account for roughly two-thirds of total spillins for firms in $Q_1$–$Q_4$, and as much as 83.1\% for $Q_5$ (the largest firms). Consistent with the findings in Section \ref{subsection:homophily_financial_characteristics}, these results point to a pronounced form of size-based homophily in the propagation of spillovers: facilities receive most of their indirect effects from other facilities operated by firms of similar size.

This pattern is economically intuitive. Technological compatibility and scale effects imply that firms of comparable size tend to employ similar production technologies and levels of capital intensity, while environmental technologies and abatement investments typically exhibit economies of scale and minimum efficient scales of adoption \citep[see, e.g.][]{jov1982,bar2000,por1995,bru2003}. In addition, benchmarking, disclosure practices, and financial and regulatory pressures are strongly stratified by firm size: large firms primarily compare environmental performance to other large, visible peers facing similar stakeholder and investor scrutiny, whereas smaller firms tend to emulate behaviour within their own size segment, under looser oversight and tighter financing constraints \citep{lea2014,fou2014,mat2014,bol2021,cla2008,mar2014,alb2019}. More generally, network formation theory predicts homophily in characteristics that reduce coordination costs and uncertainty \citep{jac2007,sor2008}, which is consistent with the size-based clustering observed in our spillin estimates.

Finally, Table~\ref{tab:spillins1} contrasts these findings with spillins obtained when the network is imposed ex ante as $W_{\text{firm}}$, $W_{\text{industry}}$, or $W_{\text{state}}$. By construction, these matrices connect each facility only to other facilities in the same group, so all estimated spillins are ``within-group''. The key comparison concerns the magnitude of these within-group spillins relative to those implied by $\widehat{\mathbf{W}}$. For instance, the industry-based network $W_{\text{industry}}$ yields an asset spillin of 0.248, whereas the within-industry component under $\widehat{\mathbf{W}}$ is only 0.014 (Table~\ref{tab:spillins}); the former is therefore about 18 times larger. The same pattern holds for sales and capex.

From a policy perspective, this discrepancy is non-trivial. A regulator or investor using an industry-based network to assess the indirect impact of sector-specific measures (e.g.\ tighter standards or engagement campaigns aimed at a given industry) would infer much stronger within-industry propagation than is implied by the data-driven network. In practice, this would lead to overstated predictions of how far an industry-focused intervention can diffuse through peer effects, and to underestimation of spillovers that operate across industries or along other dimensions such as firm size. More generally, this illustrates that a network specification which forces links to lie entirely within firms, industries, or states substantially amplifies the apparent importance of within-group spillins, whereas the empirically inferred network suggests that such group links account for only a modest share of total indirect effects.

Taken together, these comparisons highlight that the structure and strength of spillins are highly sensitive to how the network is specified: data-driven networks point to size-based patterns in link formation and extensive cross-group interactions, whereas ad hoc group networks concentrate all propagation within predefined categories and may considerably overstate group-specific spillovers.

\subsection{Robustness and Extended Specifications}\label{sec:robust}

In this section, we perform a series of robustness checks to assess whether the spillover effects identified in our benchmark specification are driven by alternative modelling choices or omitted variables. In particular, we re-estimate the model including (i) a lagged dependent variable, $y_{i,t-1}$ (lagged change in emissions), to control for emissions persistence and (ii) additional corporate governance controls.

As shown in Table~\ref{tab:robustness}, the coefficient on the lagged term is negative and highly significant ($-0.256$, p$<0.001$), indicating partial reversion or compensatory adjustment in emissions over time. Crucially, the inclusion of $y_{i,t-1}$ leaves the magnitude and significance of the spatial lag coefficient essentially unchanged, implying that network spillovers operate over and above facility-specific historical trends. This suggests that our main findings are not driven by autocorrelation in emissions behaviour.

The remaining columns of Table~\ref{tab:robustness} report results from specifications augmented with firm-level corporate governance controls, including CEO gender, board age diversity, and executive team structure. The coefficients on these variables are generally statistically insignificant, and their inclusion does not materially alter either the spatial lag coefficient or the direct effects of the financial covariates. Governance characteristics, while potentially important for other dimensions of firm behaviour, therefore do not appear to confound the relationship between financial variables and emissions spillovers in our setting. The stability of the main coefficients across these augmented specifications further supports the robustness of the network effects documented above.

\section{Conclusions}\label{sec:conclusions}

Using recent advances in the econometrics of networks, this study shows that facility-level spillover effects in emission behaviour are captured more accurately by a data-driven network than by exogenously imposed structures such as geographic proximity or industry classification.

Drawing on a panel of 399 industrial facilities, associated with 98 U.S.\ publicly listed parent firms and observed from 2000 to 2023, our heterogeneous spatial panel model provides evidence on both direct and indirect effects on emissions. Firm size, capital expenditure, and sales performance emerge as statistically significant predictors of own emissions. Indirect effects transmitted through the estimated network account for a sizeable share of the total impact of key firm characteristics, suggesting that emissions behaviour diffuses through latent corporate networks plausibly shaped by competitive dynamics, investor coalitions, supply chains, and reputational pressures.

Methodologically, the paper demonstrates the feasibility and empirical gains from relaxing the standard assumption that the network of interactions is known a priori. By combining a data-driven estimator of the adjacency matrix with a Mean Group IV approach that allows for heterogeneous slopes and interactive fixed effects, it recovers both the underlying network structure and the associated spillovers without imposing $\mathbf{W}$ ex ante.

Although our application concerns corporate emissions, the approach is applicable more broadly to settings in finance, innovation, and industrial organisation where relational dependencies are complex and only partially observed.

From a policy and practice perspective, the findings underscore that corporate emissions are shaped not only by firm-specific characteristics but also by a wider web of inter-firm interdependencies. Environmental decisions are, at least in part, strategic and relational: facilities adjust their pollution behaviour in response to the actions and characteristics of interconnected peers. This challenges models that treat firms as isolated units. For regulators and institutional investors, the presence of spillovers implies that interventions targeting groups of interconnected facilities, rather than isolated plants, may generate amplified responses through indirect channels. For analysts, the results highlight the value of flexible, data-driven measures of firm connectivity in uncovering hidden structures of influence that shape environmental and strategic outcomes.


\bibliographystyle{apalike}
\bibliography{allPapersClimate}

\begin{sidewaystable}
\centering
\caption{\textbf{Industry Category Classification Based on SIC Codes Included in the Sample}}
\begin{tabular}{cllp{5.5cm}c}
\hline
\textbf{Category} & \textbf{Name} & \textbf{SIC Code} & \textbf{Description} & \textbf{Frequency} \\
\hline
1 & Mining & 10, 13 & Metal and coal mining, oil and gas extraction & 3.2\% \\
2 & Food Products & 20 & Food and kindred product manufacturing & 10.7\% \\
3 & Wood, Furniture, Paper & 24--26 & Wood products, furniture and fictures, paper & 8.7\% \\
4 & Chemicals and Petroleum & 28--30 & Chemical production, petroleum refining, plastics & 14.5\% \\
5 & Metal Products & 33--34 & Primary metal industries and fabricated metal products & 17.0\% \\
6 & Machinery, Electronics, Instruments & 35--36, 38 & Industrial machinery, electronic equipment, scientific instruments & 13.5\% \\
7 & Transport Equipment & 37 & Motor vehicles, aircraft, and other transportation equipment & 12.2\% \\
8 & Utilities & 49 & Electric, gas, and sanitary services & 7.0\% \\
9 & Public Administration & 99 & Government agencies and public administration & 13.2\% \\
\hline
\end{tabular}
\label{tab:sic_9categories}
\end{sidewaystable}

\begin{sidewaystable}[htbp]\centering
\caption{\textbf{Variable Definitions, Data Sources and Summary Statistics}}
\label{tab:datadef}
\renewcommand\thetable{Appendix}
\def\sym#1{\ifmmode^{#1}\else\(^{#1}\)\fi}

\resizebox{\columnwidth}{!}{%
\centering
\begin{tabular}{@{}lllccc@{}}
\toprule
\hline
\textbf{Variable Group} &
  \textbf{Definition} &
  \textbf{Data source} &
  \multicolumn{3}{c}{\textbf{Summary Stats}} \\
& & & \textbf{Mean} & \textbf{Median} & \textbf{SD} \\ \midrule
\hline

\textbf{A. Dependent variable} & & & & & \\
\cline{1-1}

$\Delta$ Total Emissions $(y_{i,t})$ &
  \begin{tabular}[l]{@{}l@{}}
  Total air, water, and ground emissions (in million pounds) released \\
  on-site at the reporting facility. Water emissions consist of releases \\
  to streams and other surface bodies of water. Ground emissions \\
  consist of waste disposed in underground injection wells, landfills, \\
  surface impoundments, or spills and leaks released to land.
  \end{tabular}
  & TRI (EPA) &
  $-0.042$ & $0.000$ & $0.516$ \\

\hline

\textbf{B. Firm-level variables} & & & & & \\
\cline{1-1}

$\ln(\text{Assets}_{i,t-1})$ $(x_{1,i,t})$ &
  Value of total firm assets. &
  Compustat &
  $9.860$ & $9.656$ & $1.870$ \\

Leverage$_{i,t-1}$ $(x_{2,i,t})$ &
  (Debt in Current Liabilities + Long − Term Debt)/ Total Assets. &
  Compustat &
  $0.295$ & $0.280$ & $0.150$ \\

$\ln(\text{Capital Expenditure}_{i,t-1})$ $(x_{3,it})$ &
  Capital Expenditure / Lagged Property, Plant and Equipment. &
  Compustat &
  $6.341$ & $6.344$ & $1.975$ \\

Cash Ratio$_{i,t-1}$ $(x_{4,it})$ &
  (Cash + Cash Equivalents) / Net Assets. &
  Compustat &
  $0.090$ & $0.072$ & $0.074$ \\

$\ln(\text{Sales}_{i,t-1})$ $(x_{5,i,t})$ &
  Establishment Sales (in million US dollars). &
  Compustat &
  $9.605$ & $9.471$ & $1.615$ \\

\multirow{2}{*}{Tobin’s Q$_{i,t-1}$ $(x_{6,i,t})$} &
  (Total Assets + Common Shares Outstanding $\times$ Closing Price  &
  \multirow{2}{*}{Compustat} &
  $1.540$ & $1.302$ & $0.687$ \\
& -- Common Equity -- Deferred Taxes) / Total Assets. & & & & \\

Tangible Assets$_{i,t-1}$ $(x_{7,i,t})$ &
  Ratio of net property, plant, and equipment to total assets &
  Compustat &
  $0.344$ & $0.278$ & $0.211$ \\

\hline

\textbf{C. Corporate Governance} & & & & & \\
\cline{1-1}

Options Granted$_{i,t-1}$ $(x_{8,i,t})$ &
   Options Granted (\$ - Compustat Black Scholes value) &
   Execucomp &
   $2.910$ & $4.055$ & $2.759$ \\

Female CEO$_{i,t-1}$ $(x_{9,i,t})$ &
   Dummy variable equals 1 if the CEO is female, and 0 otherwise. &
   Execucomp &
   $0.030$ & $0.000$ & $0.170$ \\

Board Age Diversity$_{i,t-1}$ $(x_{10,i,t})$ &
   Dummy that takes value 1 for AGE\_EXECU>=60 and 0 otherwise. &
   Execucomp &
   $0.460$ & $0.000$ & $0.498$ \\

Executive Director$_{i,t-1}$ $(x_{11,i,t})$ &
  Executive served as a director during the fiscal year &
  Execucomp &
  $0.989$ & $1.000$ & $0.102$ \\

CEO Remuneration$_{i,t-1}$ $(x_{12,i,t})$ &
  Executive Rank by Salary + Bonus in the indicated fiscal year. &
  Execucomp &
  $1.236$ & $1.000$ & $0.640$ \\

\hline
\bottomrule
\end{tabular}
}

\vspace{0.5em}
\begin{minipage}{0.95\columnwidth}
\footnotesize \textit{Notes:} This table presents the definition and source of the variables in our dataset.
Panel A presents definition for the dependent variables, Panel B for firm-level control variables, and Panel C
for corporate governance control characteristics. The abbreviation TRI stands for the Toxics Release Inventory.
\end{minipage}

\end{sidewaystable}

\begin{sidewaystable}
\centering
\caption{\textbf{Summary of Network Adjacency Matrices}}
\label{tab:weights}
\begin{tabular}{llp{7.2cm}p{3.8cm}}
\hline
\textbf{Network Type} & \textbf{Definition of $w_{i,j}$} & \textbf{Notes} & \textbf{Characteristics} \\
\hline

\textbf{(1)} Estimated network, $\widehat{\boldsymbol{W}}$ &
$w_{i,j}=$$\begin{cases}
1 & \text{if } \widehat{\mathcal{J}}_{i,j}=1
 \\
0 & \text{otherwise}
\end{cases}$ &
Data-driven estimation using the method of \citet{JuodisEtAl2025}. Represents the inferred network of interactions among firms. &
Density = 0.49\%; Average links per plant = 1.95. Neighbors from the same industry/state/firm: 14.7\%/4.0\%/3.3\%.\\

\textbf{(2)} Threshold-distance &
$w_{i,j}=$$\begin{cases}
1/d_{i,j} & \text{if } d_{i,j} \leq x \text{ miles} \\
0 & \text{otherwise}
\end{cases}$ &
$d_{i,j}$ is the great-circle distance (Haversine). Thresholds ($x$) are the 10th, 5th, and 1st percentiles of the $d_{i,j}$ distribution. &
Densities are 10\%, 5\% and 1\%, respectively; Average links per plant: 40, 20, 4. \\

\textbf{(3)} k-nearest links &
$w_{i,j}=$$\begin{cases}
1 & \text{if } j \in \text{k-nearest to } i \\
0 & \text{otherwise}
\end{cases}$ &
$d_{i,j}$ as defined above. Each firm is linked to its 5 or 2 nearest neighbors by $d_{i,j}$. &
Each firm has exactly 5 or 2 links; symmetric; densities: 1.26\% and 0.50\%, respectively.\\

\textbf{(4)} Gaussian kernel &
$w_{i,j}=$$\exp\left(-\frac{d_{i,j}^{2}}{2\sigma^2}\right)$ & $d_{i,j}$ is as above.
$\sigma$ is a bandwidth parameter given by the sample standard deviation of $d_{i,j}$ divided by 3. &
All pairs receive positive weights; Most are near zero; Density = 7.54\%; Avg. links per plant = 30. \\

\textbf{(5)} By category &
$w_{i,j}=$$\begin{cases}
1 & \text{if } j \in \text{same category as } i \\
0 & \text{otherwise}
\end{cases}$ & Categories are ``industry'', ``state'' or ``firm''. &
Densities: 12.33\%, 3.47\%, 2.24\%, respectively; Average links per plant = 49, 14, 9. \\
\hline
\end{tabular}

\vspace{0.5em}
\begin{minipage}{1\linewidth}
\footnotesize \textbf{Note:} All weighting matrices are row-normalised (rows sum to one). For the data-driven network, see the Appendix for the definition of $\widehat{\mathcal{J}}_{i,j}$. For the Gaussian kernel, summary statistics use the cut-off $w_{i,j}>0.01$.
\end{minipage}
\end{sidewaystable}

\begin{sidewaystable}
\centering
\caption{\textbf{Results for Spatial MGIV Estimators with Different Network Matrices}}
\label{tab:main_results_coeff_A}
\begin{tabular}{lcccccccccc}
\hline
Variable & $\widehat{W}$ & $W_{10\%}$ & $W_{5\%}$ & $W_{1\%}$ & $W_{\text{Gaussian}}$ &  $W_{5\text{nn}}$  & $W_{2\text{nn}}$ & $W_{\text{firm}}$ & $W_{\text{industry}}$ & $W_{\text{state}}$ \\
\hline
$x_{1,i,t}$ (assets) & 0.266*** & 0.293*** & 0.329*** & 0.284*** & 0.373*** & 0.313*** & 0.312*** & 0.186* & 0.314*** & 0.281*** \\
                     & (0.099)  & (0.103)  & (0.102)  & (0.097)  & (0.110)  & (0.100)  & (0.102)  & (0.107) & (0.106)  & (0.104) \\
$x_{2,i,t}$ (leverage) & -3.794 & -0.944 & 0.265 & -0.353 & -0.518 & 2.478 & 1.033 & -2.576 & 2.019 & -1.251 \\
                     & (6.958)  & (5.850)  & (3.777)  & (3.275)  & (6.033)  & (5.648)  & (3.308)  & (5.598) & (2.517)  & (4.103) \\
$x_{3,i,t}$ (capex)  & -0.070* & -0.061 & -0.067 & -0.062 & -0.020 & -0.074* & -0.069 & -0.047 & -0.038 & -0.073* \\
                     & (0.038)  & (0.041)  & (0.041)  & (0.041)  & (0.043)  & (0.041)  & (0.042)  & (0.041) & (0.041)  & (0.042) \\
$x_{4,i,t}$ (cash)   & -0.288 & -0.134 & -0.292 & -0.483 & -0.766** & -0.510 & -0.561 & -0.004 & -0.303 & -0.057 \\
                     & (0.405)  & (0.401)  & (0.388)  & (0.375)  & (0.355)  & (0.397)  & (0.373)  & (0.429) & (0.396)  & (0.415) \\
$x_{5,i,t}$ (sales)  & -0.287*** & -0.261** & -0.252** & -0.264*** & -0.383*** & -0.271*** & -0.311*** & -0.194 & -0.277** & -0.262** \\
                     & (0.097)  & (0.109)  & (0.104)  & (0.097)  & (0.112)  & (0.102)  & (0.102)  & (0.120) & (0.109)  & (0.105) \\
$x_{6,i,t}$ (Tobin's Q) & 0.059 & 0.095 & 0.096 & 0.107 & 0.045 & 0.063 & 0.095 & 0.032 & 0.021 & 0.092 \\
                     & (0.061)  & (0.075)  & (0.076)  & (0.076)  & (0.072)  & (0.077)  & (0.078)  & (0.075) & (0.071)  & (0.073) \\
$x_{7,i,t}$ (tangibles) & -0.170 & 0.162 & 0.200 & -0.039 & -0.150 & -0.034 & 0.075 & -0.101 & -0.078 & -0.184 \\
                     & (0.352)  & (0.354)  & (0.365)  & (0.350)  & (0.333)  & (0.346)  & (0.354)  & (0.323) & (0.324)  & (0.330) \\
\hline
$W_{y}$ (spatial lag) & 0.285*** & 0.091 & 0.086 & 0.063** & 0.148* & 0.016** & 0.023 & 0.395*** & 0.444*** & 0.156*** \\
                      & (0.046)  & (0.072) & (0.058) & (0.031) & (0.085) & (0.008) & (0.015) & (0.047)  & (0.090)  & (0.057) \\
\hline
$N$ & 399 & 399 & 399 & 399 & 399 & 399 & 399 & 399 & 399 & 399 \\
$r$ & 2 & 2 & 2 & 2 & 2 & 2 & 2 & 2 & 2 & 2 \\
\hline
\end{tabular}
\vspace{0.5em}
\parbox{\linewidth}{
\small
\textbf{Notes:} $\widehat{W}$ denotes the network matrix estimated from the data. $W_{10\%}$, $W_{5\%}$, and $W_{1\%}$ are constructed using great-circle distance thresholds at the 10th, 5th, and 1st percentiles, respectively. $W_{\text{Gaussian}}$ employs a Gaussian kernel for great-circle distance decay. $W_{5\text{nn}}$ and $W_{2\text{nn}}$ are based on the 5 and 2 nearest neighbors, respectively (by great-circle distance). Finally, $W_{\text{firm}}$, $W_{\text{industry}}$ and $W_{\text{state}}$ connect each $i$ to all $j$ in the same category (either firm, industry or state). Standard errors in parentheses. \quad * $p<0.10$, ** $p<0.05$, *** $p<0.01$.
}
\end{sidewaystable}
\vspace*{\fill}

\begin{sidewaystable}
\centering
\caption{\textbf{Results for Non-Spatial Estimators}}
\label{tab:main_results_coeff_B}
\begin{tabular}{lcccc}
\hline
Variable & TWFE-firm & TWFE-facility & 2SIV & MGIV \\
\hline
$x_{1,i,t}$ (assets) & -0.026 & -0.009 & 0.052 & 0.331*** \\
                     & (0.026) & (0.023) & (0.034) & (0.099) \\
$x_{2,i,t}$ (leverage) & 0.090 & 0.060 & 0.049 & -0.522 \\
                     & (0.069) & (0.068) & (0.087) & (3.099) \\
$x_{3,i,t}$ (capex)  & 0.005 & 0.004 & -0.028* & -0.070* \\
                     & (0.007) & (0.007) & (0.015) & (0.040) \\
$x_{4,i,t}$ (cash)   & 0.056 & 0.068 & 0.294** & -0.225 \\
                     & (0.113) & (0.110) & (0.116) & (0.406) \\
$x_{5,i,t}$ (sales)  & 0.014 & 0.007 & -0.063* & -0.254*** \\
                     & (0.029) & (0.028) & (0.037) & (0.097) \\
$x_{6,i,t}$ (Tobin's Q) & 0.005 & 0.007 & 0.010 & 0.097 \\
                     & (0.015) & (0.015) & (0.018) & (0.070) \\
$x_{7,i,t}$ (tangibles) & -0.031 & -0.031 & 0.135 & 0.011 \\
                     & (0.071) & (0.070) & (0.104) & (0.339) \\
\hline
$N$   & 399 & 399 & 399 & 399 \\
$r_x$ &  &  & 2 & 2 \\
$r_y$ &  &  & 1 & \\
\hline
\end{tabular}

\vspace{0.5em}
\begin{minipage}{1\linewidth}
\footnotesize \textbf{Notes:} “TWFE-firm/facility” are two-way fixed-effects estimators with common time effects and firm- or facility-specific fixed effects, respectively. “2SIV” denotes the two-stage IV estimator of \citet{NorkuteEtal2021} and \citet{CuiEtal2022}. “MGIV” is the mean-group IV estimator of \citet{NorkuteEtal2021}. None of these estimators incorporate a spatial network specification. Standard errors in parentheses. \quad * $p<0.10$, ** $p<0.05$, *** $p<0.01$.
\end{minipage}
\end{sidewaystable}
\vspace*{\fill}

\begin{sidewaystable}
\centering
\caption{\textbf{Impact Decomposition for Spatial MGIV Estimators - Direct Effects}}
\label{tab:impact_decomposition_DE}
\begin{tabular}{lcccccccccc}
\hline
Variable & $\widehat{W}$ & $W_{10\%}$ & $W_{5\%}$ & $W_{1\%}$ & $W_{\text{Gaussian}}$ &  $W_{5\text{nn}}$  & $W_{2\text{nn}}$ & $W_{\text{firm}}$ & $W_{\text{industry}}$ & $W_{\text{state}}$ \\
\hline
$x_{1,i,t}$ (assets) & 0.267*** & 0.293*** & 0.329*** & 0.284*** & 0.373*** & 0.313*** & 0.312*** & 0.195* & 0.316*** & 0.282*** \\
                     & (0.099)  & (0.103)  & (0.102)  & (0.097)  & (0.110)  & (0.100)  & (0.102)  & (0.113) & (0.106)  & (0.104) \\
$x_{2,i,t}$ (leverage) & -3.806 & -0.945 & 0.265 & -0.353 & -0.518 & 2.480 & 1.034 & -2.704 & 2.035 & -1.256 \\
                     & (6.980)  & (5.854)  & (3.782)  & (3.281)  & (6.039)  & (5.654)  & (3.310)  & (5.879) & (2.538)  & (4.117) \\
$x_{3,i,t}$ (capex)  & -0.071* & -0.061 & -0.067 & -0.062 & -0.020 & -0.074* & -0.069 & -0.049 & -0.038 & -0.074* \\
                     & (0.039)  & (0.041)  & (0.041)  & (0.041)  & (0.043)  & (0.041)  & (0.042)  & (0.043) & (0.041)  & (0.042) \\
$x_{4,i,t}$ (cash)   & -0.289 & -0.134 & -0.293 & -0.484 & -0.767** & -0.511 & -0.561 & -0.004 & -0.305 & -0.057 \\
                     & (0.406)  & (0.402)  & (0.388)  & (0.376)  & (0.355)  & (0.398)  & (0.374)  & (0.450) & (0.399)  & (0.416) \\
$x_{5,i,t}$ (sales)  & -0.288*** & -0.261** & -0.253** & -0.264*** & -0.383*** & -0.271*** & -0.311*** & -0.204 & -0.279** & -0.263** \\
                     & (0.097)  & (0.109)  & (0.104)  & (0.097)  & (0.112)  & (0.102)  & (0.102)  & (0.126) & (0.110)  & (0.105) \\
$x_{6,i,t}$ (Tobin's Q) & 0.059 & 0.095 & 0.096 & 0.107 & 0.045 & 0.063 & 0.095 & 0.034 & 0.022 & 0.092 \\
                     & (0.061)  & (0.075)  & (0.076)  & (0.076)  & (0.072)  & (0.077)  & (0.078)  & (0.078) & (0.072)  & (0.073) \\
$x_{7,i,t}$ (tangibles) & -0.170 & 0.162 & 0.201 & -0.039 & -0.151 & -0.034 & 0.075 & -0.106 & -0.078 & -0.185 \\
                     & (0.353)  & (0.355)  & (0.365)  & (0.350)  & (0.333)  & (0.347)  & (0.354)  & (0.339) & (0.327)  & (0.331) \\
\hline
\end{tabular}
\vspace{0.5em}
\parbox{\linewidth}{
\small
\textbf{Notes:} $\widehat{W}$ denotes the network matrix estimated from the data. $W_{10\%}$, $W_{5\%}$, and $W_{1\%}$ are constructed using great-circle distance thresholds at the 10th, 5th, and 1st percentiles, respectively. $W_{\text{Gaussian}}$ employs a Gaussian kernel for great-circle distance decay. $W_{5\text{nn}}$ and $W_{2\text{nn}}$ are based on the 5 and 2 nearest neighbors, respectively (by great-circle distance). Finally, $W_{\text{firm}}$, $W_{\text{industry}}$ and $W_{\text{state}}$ connect each $i$ to all $j$ in the same category (either firm, industry, or state). Standard errors in parentheses. \quad * $p<0.10$, ** $p<0.05$, *** $p<0.01$.
}
\end{sidewaystable}
\vspace*{\fill}

\begin{sidewaystable}
\centering
\caption{\textbf{Impact Decomposition for Spatial MGIV Estimators - Indirect Effects}}
\label{tab:impact_decomposition_IE}
\begin{tabular}{lcccccccccc}
\hline
Variable & $\widehat{W}$ & $W_{10\%}$ & $W_{5\%}$ & $W_{1\%}$ & $W_{\text{Gaussian}}$ &  $W_{5\text{nn}}$  & $W_{2\text{nn}}$ & $W_{\text{firm}}$ & $W_{\text{industry}}$ & $W_{\text{state}}$ \\
\hline
$x_{1,i,t}$ (assets) & 0.105** & 0.029 & 0.031 & 0.018 & 0.065 & 0.027 & 0.015 & 0.102* & 0.248** & 0.050* \\
                     & (0.043) & (0.026) & (0.024) & (0.012) & (0.045) & (0.017) & (0.011) & (0.061) & (0.117) & (0.028) \\
$x_{2,i,t}$ (leverage) & -1.501 & -0.094 & 0.025 & -0.023 & -0.090 & 0.217 & 0.049 & -1.416 & 1.597 & -0.224 \\
                     & (2.704) & (0.577) & (0.353) & (0.213) & (1.041) & (0.515) & (0.160) & (3.100) & (2.076) & (0.733) \\
$x_{3,i,t}$ (capex)  & -0.028* & -0.006 & -0.006 & -0.004 & -0.003 & -0.006 & -0.003 & -0.026 & -0.030 & -0.013 \\
                     & (0.016) & (0.007) & (0.006) & (0.003) & (0.008) & (0.005) & (0.003) & (0.023) & (0.033) & (0.009) \\
$x_{4,i,t}$ (cash)   & -0.114 & -0.013 & -0.027 & -0.031 & -0.133 & -0.045 & -0.027 & -0.002 & -0.240 & -0.010 \\
                     & (0.163) & (0.041) & (0.040) & (0.030) & (0.107) & (0.041) & (0.026) & (0.236) & (0.316) & (0.075) \\
$x_{5,i,t}$ (sales)  & -0.114*** & -0.026 & -0.024 & -0.017 & -0.066 & -0.024 & -0.015 & -0.107 & -0.219* & -0.047* \\
                     & (0.042) & (0.023) & (0.018) & (0.011) & (0.045) & (0.015) & (0.011) & (0.069) & (0.118) & (0.027) \\
$x_{6,i,t}$ (Tobin's Q) & 0.023 & 0.009 & 0.009 & 0.007 & 0.008 & 0.006 & 0.004 & 0.018 & 0.017 & 0.016 \\
                     & (0.025) & (0.011) & (0.010) & (0.006) & (0.014) & (0.008) & (0.005) & (0.041) & (0.056) & (0.015) \\
$x_{7,i,t}$ (tangibles) & -0.067 & 0.016 & 0.019 & -0.003 & -0.026 & -0.003 & 0.004 & -0.056 & -0.061 & -0.033 \\
                     & (0.139) & (0.039) & (0.037) & (0.023) & (0.058) & (0.030) & (0.017) & (0.177) & (0.257) & (0.061) \\
\hline
\end{tabular}
\vspace{0.5em}
\parbox{\linewidth}{
\small
\textbf{Notes:} $\widehat{W}$ denotes the network matrix estimated from the data. $W_{10\%}$, $W_{5\%}$, and $W_{1\%}$ are constructed using great-circle distance thresholds at the 10th, 5th, and 1st percentiles, respectively. $W_{\text{Gaussian}}$ employs a Gaussian kernel for great-circle distance decay. $W_{5\text{nn}}$ and $W_{2\text{nn}}$ are based on the 5 and 2 nearest neighbors, respectively (by great-circle distance). Finally, $W_{\text{firm}}$, $W_{\text{industry}}$ and $W_{\text{state}}$ connect each $i$ to all $j$ in the same category (either firm, industry, or state). Standard errors in parentheses. \quad * $p<0.10$, ** $p<0.05$, *** $p<0.01$.
}
\end{sidewaystable}

\begin{sidewaystable}
\centering
\caption{\textbf{Impact Decomposition for Spatial MGIV Estimators - Total Effects}}
\label{tab:impact_decomposition_TE}
\begin{tabular}{lcccccccccc}
\hline
Variable & $\widehat{W}$ & $W_{10\%}$ & $W_{5\%}$ & $W_{1\%}$ & $W_{\text{Gaussian}}$ &  $W_{5\text{nn}}$  & $W_{2\text{nn}}$ & $W_{\text{firm}}$ & $W_{\text{industry}}$ & $W_{\text{state}}$ \\
\hline
$x_{1,i,t}$ (assets) & 0.372*** & 0.322*** & 0.360*** & 0.303*** & 0.438*** & 0.340*** & 0.327*** & 0.297* & 0.565*** & 0.332*** \\
                     & (0.137)  & (0.114)  & (0.112)  & (0.105)  & (0.129)  & (0.110)  & (0.108)  & (0.172) & (0.204)  & (0.125) \\
$x_{2,i,t}$ (leverage) & -5.308 & -1.039 & 0.290 & -0.376 & -0.608 & 2.697 & 1.083 & -4.120 & 3.633 & -1.480 \\
                     & (9.670)  & (6.425)  & (4.134)  & (3.493)  & (7.078)  & (6.158)  & (3.467)  & (8.972) & (4.569)  & (4.845) \\
$x_{3,i,t}$ (capex)  & -0.098* & -0.067 & -0.073 & -0.066 & -0.023 & -0.081* & -0.072 & -0.074 & -0.069 & -0.087* \\
                     & (0.054)  & (0.046)  & (0.046)  & (0.044)  & (0.050)  & (0.044)  & (0.044)  & (0.066) & (0.073)  & (0.050) \\
$x_{4,i,t}$ (cash)   & -0.402 & -0.148 & -0.320 & -0.515 & -0.900** & -0.555 & -0.588 & -0.007 & -0.545 & -0.067 \\
                     & (0.567)  & (0.442)  & (0.424)  & (0.401)  & (0.424)  & (0.433)  & (0.392)  & (0.686) & (0.708)  & (0.491) \\
$x_{5,i,t}$ (sales)  & -0.402*** & -0.287** & -0.276** & -0.282*** & -0.450*** & -0.295*** & -0.326*** & -0.311 & -0.499** & -0.310** \\
                     & (0.134)   & (0.118)  & (0.112)  & (0.104)   & (0.130)   & (0.111)  & (0.107)  & (0.194) & (0.213)  & (0.125) \\
$x_{6,i,t}$ (Tobin's Q) & 0.082 & 0.105 & 0.105 & 0.114 & 0.053 & 0.069 & 0.100 & 0.051 & 0.039 & 0.108 \\
                     & (0.086)   & (0.082)  & (0.084)  & (0.082)  & (0.085)  & (0.084)  & (0.082)  & (0.119) & (0.128)  & (0.087) \\
$x_{7,i,t}$ (tangibles) & -0.238 & 0.178 & 0.219 & -0.042 & -0.177 & -0.037 & 0.079 & -0.162 & -0.140 & -0.218 \\
                     & (0.491)  & (0.391)  & (0.400)  & (0.373)  & (0.389)  & (0.377)  & (0.371)  & (0.517) & (0.584)  & (0.391) \\
\hline
\end{tabular}
\vspace{0.5em}
\parbox{\linewidth}{
\small
\textbf{Notes:} $\widehat{W}$ denotes the network matrix estimated from the data. $W_{10\%}$, $W_{5\%}$, and $W_{1\%}$ are constructed using great-circle distance thresholds at the 10th, 5th, and 1st percentiles, respectively. $W_{\text{Gaussian}}$ employs a Gaussian kernel for great-circle distance decay. $W_{5\text{nn}}$ and $W_{2\text{nn}}$ are based on the 5 and 2 nearest neighbors, respectively (by great-circle distance). Finally, $W_{\text{firm}}$, $W_{\text{industry}}$ and $W_{\text{state}}$ connect each $i$ to all $j$ in the same category (either firm, industry or state). Standard errors in parentheses. \quad * $p<0.10$, ** $p<0.05$, *** $p<0.01$.
}
\end{sidewaystable}

\begin{sidewaystable}
\begin{table}[H]
\centering
\caption{\textbf{Spillins for the Spatial MGIV Estimator based on} $\widehat{\mathbf{W}}$\textbf{, by firm, industry and state.}}
\label{tab:spillins}
\begin{tabular}{l|ccc|ccc|ccc}
\hline
 & \multicolumn{3}{c|}{\text{by firm}} & \multicolumn{3}{c|}{\text{by industry}} & \multicolumn{3}{c}{\text{by state}} \\
\cline{2-4}\cline{5-7}\cline{8-10}
 & \text{Within} & \text{Between} & \text{All} & \text{Within} & \text{Between} & \text{All} & \text{Within} & \text{Between} & \text{All} \\
\hline
$x_{1,i,t}$ (assets) & 0.002 & 0.103 & 0.105 & 0.014 & 0.092 & 0.105 & 0.004 & 0.101 & 0.105 \\
\textit{share}       & (2.2\%) & (97.8\%) & (100\%) & (13.1\%) & (86.9\%) & (100\%) & (3.9\%) & (96.1\%) & (100\%) \\
\hline
$x_{3,i,t}$ (capex)  & -0.001 & -0.027 & -0.028 & -0.004 & -0.024 & -0.028 & -0.001 & -0.027 & -0.028 \\
\textit{share}       & (2.2\%) & (97.8\%) & (100\%) & (13.1\%) & (86.9\%) & (100\%) & (3.9\%) & (96.1\%) & (100\%) \\
\hline
$x_{5,i,t}$ (sales)  & -0.002 & -0.111 & -0.114 & -0.015 & -0.099 & -0.114 & -0.004 & -0.109 & -0.114 \\
\textit{share}       & (2.2\%) & (97.8\%) & (100\%) & (13.1\%) & (86.9\%) & (100\%) & (3.9\%) & (96.1\%) & (100\%) \\
\hline
\end{tabular}

\vspace{0.5em}
\footnotesize\emph{Notes:} $\widehat{W}$ denotes the network matrix estimated from the data. ``Within'' reports indirect effects from same-firm, same-industry or same-state links. ``Between'' reports indirect effects from different-firm, different-industry or different-state links. Percentages (as shares of total spillins) are shown within brackets beneath the Within/Between entries. The “All” column entries match the corresponding indirect effects in Table~\ref{tab:impact_decomposition_IE}.
\end{table}
\end{sidewaystable}

\begin{sidewaystable}
\centering
\caption{\textbf{Spillins for the Spatial MGIV Estimator based on} $\widehat{\mathbf{W}}$\textbf{, for different quintiles of the distribution of firm assets.}}
\label{tab:spillins_quintile}
\begin{tabular}{l|cc|cc|cc|cc|cc|c}
\hline
 & \multicolumn{2}{c|}{Q1} & \multicolumn{2}{c|}{Q2} & \multicolumn{2}{c|}{Q3} & \multicolumn{2}{c|}{Q4} & \multicolumn{2}{c|}{Q5} & \multicolumn{1}{c}{All} \\
\cline{2-3}\cline{4-5}\cline{6-7}\cline{8-9}\cline{10-11}\cline{12-12}
 & Within & Between & Within & Between & Within & Between & Within & Between & Within & Between &  \\
\hline
$x_{1,i,t}$ (assets) & 0.072 & 0.034 & 0.075 & 0.030 & 0.072 & 0.034 & 0.069 & 0.037 & 0.088 & 0.018 & 0.105 \\
\textit{share}       & (68.1\%) & (31.9\%) & (71.1\%) & (28.9\%) & (68.0\%) & (32.0\%) & (65.3\%) & (34.7\%) & (83.1\%) & (16.9\%) & (100.0\%) \\
\hline
$x_{3,i,t}$ (capex)  & -0.019 & -0.009 & -0.020 & -0.008 & -0.019 & -0.009 & -0.018 & -0.010 & -0.023 & -0.005 & -0.028 \\
\textit{share}       & (68.1\%) & (31.9\%) & (71.1\%) & (28.9\%) & (68.0\%) & (32.0\%) & (65.3\%) & (34.7\%) & (83.1\%) & (16.9\%) & (100.0\%) \\
\hline
$x_{5,i,t}$ (sales)  & -0.077 & -0.036 & -0.081 & -0.033 & -0.077 & -0.036 & -0.074 & -0.039 & -0.094 & -0.019 & -0.114 \\
\textit{share}       & (68.1\%) & (31.9\%) & (71.1\%) & (28.9\%) & (68.0\%) & (32.0\%) & (65.3\%) & (34.7\%) & (83.1\%) & (16.9\%) & (100.0\%) \\
\hline
\end{tabular}

\vspace{0.5em}
\footnotesize\emph{Notes:} $\widehat{W}$ denotes the network matrix estimated from the data. Let $Q_j$ denote the $j$th quintile of the cross-sectional distribution of firms’ time-averaged assets. For a given quintile $Q_j$, “Within” reports the indirect effect arising from links within the same quintile ($Q_j$), whereas “Between” reports the indirect effect arising from links in different quintiles (i.e., $\bigcup_{j' \neq j} Q_{j'}$). The percentage shown beneath each pair is
$\text{Within}/(\text{Within}+\text{Between})\times 100$, i.e., the share of the quintile-specific spillin attributable to same-quintile links. For example, for $Q_5$ (largest firms), the within and between effects are $0.088$ and $0.018$, so $0.088/(0.088+0.018)=83.1\%$; thus spillins for large firms arise predominantly from links to other large firms. Similarly, for $Q_1$ (smallest firms), $68.1\%$ of the total indirect effect comes from links within $Q_1$, i.e., from other small firms. The “All” column reports the overall indirect effect, matching the corresponding indirect effects in Table~\ref{tab:impact_decomposition_IE}.
\end{sidewaystable}

\begin{sidewaystable}
\begin{table}[H]
\centering
\caption{\textbf{Spillins for Spatial MGIV Estimators based on} $W_{\text{firm}}$, $W_{\text{industry}}$ \textbf{and} $W_{\text{state}}$.}
\label{tab:spillins1}
\begin{tabular}{l|ccc|ccc|ccc}
\hline
 & \multicolumn{3}{c|}{$W_{\text{firm}}$} & \multicolumn{3}{c|}{$W_{\text{industry}}$} & \multicolumn{3}{c}{$W_{\text{state}}$} \\
\cline{2-4}\cline{5-7}\cline{8-10}
 & \text{Within} & \text{Between} & \text{All} & \text{Within} & \text{Between} & \text{All} & \text{Within} & \text{Between} & \text{All} \\
\hline
$x_{1,i,t}$ (assets) & 0.102 & 0.000 & 0.102 & 0.248 & 0.000 & 0.248 & 0.050 & 0.000 & 0.050 \\
\textit{share}       & (100.0\%) & (0.0\%) & (100\%) & (100.0\%) & (0.0\%) & (100\%) & (100.0\%) & (0.0\%) & (100\%) \\
\hline
$x_{3,i,t}$ (capex)  & -0.026 & 0.000 & -0.026 & -0.030 & 0.000 & -0.030 & -0.013 & 0.000 & -0.013 \\
\textit{share}       & (100.0\%) & (0.0\%) & (100\%) & (100.0\%) & (0.0\%) & (100\%) & (100.0\%) & (0.0\%) & (100\%) \\
\hline
$x_{5,i,t}$ (sales)  & -0.107 & 0.000 & -0.107 & -0.219 & 0.000 & -0.219 & -0.047 & 0.000 & -0.047 \\
\textit{share}       & (100.0\%) & (0.0\%) & (100\%) & (100.0\%) & (0.0\%) & (100\%) & (100.0\%) & (0.0\%) & (100\%) \\
\hline
\end{tabular}

\vspace{0.5em}
\footnotesize\emph{Notes:} $W_{\text{firm}}$, $W_{\text{industry}}$ and $W_{\text{state}}$ connect each $i$ to all $j$ in the same category (either firm, industry or state). “Within”/“Between” denote within- vs. between-group spillins for the stated grouping. Percentages (as shares of total spillins) are shown within brackets beneath the Within/Between entries. The “All” column entries match the corresponding indirect effects in Table~\ref{tab:impact_decomposition_IE}.
\end{table}
\end{sidewaystable}

\begin{table}
\centering
\caption{\textbf{Robustness Checks for Spatial MGIV based on} $\widehat{\mathbf{W}}$.}
\label{tab:robustness}
\begin{tabular}{lcccc}
\hline
\textbf{Variable} & \textbf{Baseline} & \textbf{RC1} & \textbf{RC2} & \textbf{RC3} \\
\hline
$y_{i,t-1}$ (lagged change in emissions) &            & -0.256*** &            &            \\
                                         &            & (0.018)   &            &            \\
$x_{1,it}$ (assets)                      & 0.266***   & 0.278**   & 0.326***   & 0.257**    \\
                                         & (0.099)    & (0.119)   & (0.098)    & (0.108)    \\
$x_{2,it}$ (leverage)                    & -3.794     & -4.165    & -1.857     & 5.447      \\
                                         & (6.958)    & (7.398)   & (6.180)    & (5.293)    \\
$x_{3,it}$ (capex)                       & -0.070*    & -0.082*   & -0.066*    & -0.037     \\
                                         & (0.038)    & (0.045)   & (0.040)    & (0.044)    \\
$x_{4,it}$ (cash)                        & -0.288     & -0.342    & -0.405     & -0.305     \\
                                         & (0.405)    & (0.344)   & (0.431)    & (0.558)    \\
$x_{5,it}$ (sales)                       & -0.287***  & -0.244**  & -0.380***  & -0.426***  \\
                                         & (0.097)    & (0.108)   & (0.097)    & (0.103)    \\
$x_{6,it}$ (Tobin’s Q)                   & 0.059      & 0.063     & 0.028      & 0.002      \\
                                         & (0.061)    & (0.062)   & (0.066)    & (0.070)    \\
$x_{7,it}$ (tangibles)                   & -0.170     & -0.521    & -0.043     & 0.237      \\
                                         & (0.352)    & (0.457)   & (0.351)    & (0.373)    \\
$x_{8,it}$ (options granted)             &            &           & 0.003      & 0.001      \\
                                         &            &           & (0.016)    & (0.016)    \\
$x_{9,it}$ (female CEO)                  &            &           & -0.014     & -0.021     \\
                                         &            &           & (0.012)    & (0.014)    \\
$x_{11,it}$ (executive director)         &            &           & -0.001     & 0.005      \\
                                         &            &           & (0.008)    & (0.008)    \\
$x_{10,it}$ (board age diversity)        &            &           &            & -0.022     \\
                                         &            &           &            & (0.031)    \\
$x_{12,it}$ (CEO remuneration)           &            &           &            & -0.027     \\
                                         &            &           &            & (0.024)    \\
\hline
$W_{y}$ (spatial lag of emissions)       & 0.285***   & 0.261***  & 0.283***   & 0.275***   \\
                                         & (0.046)    & (0.045)   & (0.045)    & (0.045)    \\
\hline
\end{tabular}

\vspace{0.5em}
\footnotesize
\textbf{Notes:} \textit{Baseline} corresponds to the $\widehat{W}$ specification in Table~\ref{tab:main_results_coeff_A}. \textit{RC1} adds a lagged dependent variable (lagged change in emissions). \textit{RC2}--\textit{RC3} include corporate-governance controls (two alternative sets). Standard errors in parentheses. \; * $p<0.10$, ** $p<0.05$, *** $p<0.01$.
\end{table}

\clearpage

\section*{Appendix}
\label{Appendix}

\setcounter{equation}{0}
\renewcommand{\theequation}{A.\arabic{equation}}

In this Appendix we provide further theoretical details for the estimation algorithm. We provide a formal description of the general algorithm steps while abstracting from specific details, such as the defactoring steps used in the main paper, that do not affect the algorithm implementation. We consider the
algorithm for unit $y_{i,t}$ since it is the same algorithm applied to each $%
y_{i,t}$, $i=1,...,N$, separately, until the final panel estimation step. It
is a multi-stage procedure. In the first stage, we consider the $N-1$
IV regressions of $y_{i,t}$ on $\boldsymbol{x}_{i,t}$ and $y_{j,t}$, for $%
j=1,...,i-1,i+1,...,N$. We denote the $t$-ratio $t_{i,j}$, from this
regression as $t_{i,j,(1)}$, where $(1)$ denotes stage. We select the one
with maximum absolute value of $t_{i,j,(1)}$ over $j$. We denote the index of
the selected unit by $\mathcal{S}_{i,\left( 1\right) }^{c}$. and the $%
T\times1$ observation matrix of the selected $y$ variable by $\boldsymbol{x}%
_{(i,1)}^{o}$. Further, let $\boldsymbol{X}_{i,(1)}=(\boldsymbol{X}_{i},%
\boldsymbol{x}_{(i,1)}^{o})=(\boldsymbol{x}_{(1),i,1},...,\boldsymbol{x}%
_{(1),i,T})^{\prime }$, $\mathcal{S}_{i,\left( 1\right) }=\mathcal{S}%
_{i,\left( 1\right) }^{c}$, and $\mathfrak{A}_{i,\left( 2\right) }=\left\{
1,...,i-1,i+1,...,N\right\} \setminus\mathcal{S}_{i,\left( 1\right) }$.  We
then proceed to the second stage and run regressions of $y_{i,t}$ on $%
\boldsymbol{x}_{(1),i,t}$ and $y_{j,t}$, for $j\in\mathfrak{A}_{i,\left(
2\right) }$, where appropriate instrumentation, for the included $y_{j,t},$
as well as for the selected $y$ variable in $\boldsymbol{x}_{(i,1)}^{o}$, is
carried out. We obtain the $t$-ratios for $y_{j,t}$, for $j\in\mathfrak{A}%
_{i,\left( 2\right) }$. These $t$-ratios are denoted by $t_{i,j,(2)}$. We
construct selection indicators, given by
\begin{equation}
\widehat{\mathcal{J}}_{i,j,(2)}=I[|t_{i,j,(2)}|>c_{p}\left( n,\delta\right) ]%
\text{, for }j\in\mathfrak{A}_{i,\left( 2\right) }\text{, }
\end{equation}
where $c_{p}(n,\delta)$ is the critical value function, defined by%
\begin{equation}
c_{p}\left( n,\delta\right) =\Phi^{-1}\left( 1-\frac{p}{2f\left(
n,\delta\right) }\right) \text{,}  \label{cvf1}
\end{equation}
$\Phi^{-1}\left( .\right) $ is the inverse of standard normal distribution
function, $f\left( n,\delta\right) =cn^{\delta}$ for some positive constants
$\delta$ and $c$, and $p$ ($0<p<1$) is the nominal size of the individual
tests to be set by the investigator. As in the OCMT algorithm, we will refer
to $\delta$ as the critical value exponent. Among units with $\widehat{%
\mathcal{J}}_{i,j,(2)}=1$, we select the one with maximum absolute value of $%
t_{i,j,(2)}$. We denote the index of the selected unit by $\mathcal{S}%
_{i,\left( 2\right) }^{c}$. and the $T\times1$ observation matrix of the
selected variable by $\boldsymbol{x}_{(i,2)}^{o}$. As before we let $%
\boldsymbol{X}_{i,(2)}=(\boldsymbol{X}_{i,(1)},\boldsymbol{x}_{(i,2)}^{o})=(%
\boldsymbol{x}_{(2),i,1},...,\boldsymbol{x}_{(2),i,T})^{\prime}$, $\mathcal{S%
}_{i,\left( 2\right) }=\mathcal{S}_{i,\left( 1\right) }\cup\mathcal{S}%
_{i,\left( 2\right) }^{c}$, and $\mathfrak{A}_{i,\left( 3\right) }=\left\{
1,...,i-1,i+1,...,N\right\} \setminus\mathcal{S}_{i,\left( 2\right) }$.

\bigskip We proceed to further stages, $\ell=3,...,$ in similar fashion
defining analogously $\boldsymbol{X}_{i,(\ell)}$, $\mathcal{S}_{i,\left(
\ell\right) }$ and $\mathfrak{A}_{i,(\ell+1)}$, until $\widehat{\mathcal{J}}%
_{i,j,(\ell)}=0$, for $j\in\mathfrak{A}_{i,\left( \ell\right) }$. Denote the
stage for which this happens for unit $i$ by $\hat{\ell}_{i}$. Further, we
define $\widehat{\mathcal{J}}_{i,j}=I[j\in\mathcal{S}_{i,\left( \hat{\ell }%
_{i}\right) }]$, $\mathcal{S}_{i}^{n}$ and $\mathcal{S}_{i}^{d}$ to be the
set of true $n$-units and $d$-units, for unit $i$, and
\begin{align}
\mathcal{A}_{0} & =\left\{ \cap_{i=1}^{N}\left\{ \sum \nolimits_{j\in%
\mathcal{S}_{i}^{n}}\widehat{\mathcal{J}}_{i,j}=k_{i}\right\} \right\}
\cap\left\{ \cap _{i=1}^{N}\left\{ \sum \nolimits_{j\in\mathcal{S}_{i}^{d}}%
\widehat{\mathcal{J}}_{i,j}=0\right\} \right\} \nonumber \\
& =\cap_{i=1}^{N}\left\{ \left\{ \sum \nolimits_{j\in\mathcal{S}_{i}^{n}}%
\widehat{\mathcal{J}}_{i,j}=k_{i}\right\} \cap\left\{ \sum \nolimits_{j\in%
\mathcal{S}_{i}^{d}}\widehat{\mathcal{J}}_{i,j}=0\right\} \right\}
=\cap_{i=1}^{N}\mathcal{A}_{0i}\text{.}
\end{align}
Note that for each stage, $n$ in $c_{p}\left( n,\delta\right) $ should be
set equal to the cardinality of $\mathfrak{A}_{i,\left( \ell\right) }$.

Finally, below we provide a list of formal assumptions needed for the validity of the proposition presented in the main paper.

\begin{assumption}
\label{ass1} The error terms, $\varepsilon_{i,t}$, are independent over $i$
and each is a martingale difference process with respect to $\mathcal{F}%
_{t-1}^{\varepsilon_{i}}=\sigma\left(  \varepsilon_{i,t-1},\varepsilon
_{i,t-2},...,\right)  $, with a zero mean and a constant variance,
$0<\sigma_{i}^{2}<C<\infty$.\vspace{-0.08in}
\end{assumption}

\begin{assumption}
\label{ass1A} $y_{i,t}$ and $x_{j,i,t}$, $j=1,...,p,$ $i=1,...,N$, are
strongly mixing processes with mixing coefficients given by $\alpha
_{ik}=C_{ik}\xi^{k}$ for some $C_{ik}$ such that $\sup_{i,k}C_{ik}<\infty$ and
some $0<\xi<1$. $E\left[  x_{j,i,t}\varepsilon_{i,t}\left\vert \mathcal{F}%
_{t-1}\right.  \right]  =0$, for $i=1,2,...,n,$ and all $t.$ \ \ .\vspace
{-0.08in}
\end{assumption}

\begin{assumption}
\label{ass2} There exist sufficiently large positive constants $C_{0}%
,C_{1},C_{2}$ and $C_{3}$ and $s_{x},s_{u}>0$ such that
\begin{equation}
\sup\nolimits_{j,i,t}\Pr\left(  \left\vert x_{j,i,t}\right\vert >\alpha
\right)  \leq C_{0}\exp\left(  -C_{1}\alpha^{s_{x}}\right)  ,\text{ for all
}\alpha>0,\label{expprob}%
\end{equation}
and the errors, $\varepsilon_{i,t}$, satisfy
\begin{equation}
\sup\nolimits_{i,t}\Pr\left(  \left\vert \varepsilon_{i,t}\right\vert
>\alpha\right)  \leq C_{2}\exp\left(  -C_{3}\alpha^{s_{u}}\right)  ,\text{ for
all }\alpha>0\text{.}\label{expprob2}%
\end{equation}

\end{assumption}

\begin{remark}
Assumption \ref{ass1} is standard in the literature and simply restricts the error terms to be  martingale difference processes.
Assumption \ref{ass1A} is a relaxation of much of the existing literature, such as \cite{chu2018}, since it allows mixing and therefore dynamic regressors, unlike the regressor martingale difference assumption of \cite{chu2018}. This is enabled by the sharp mixing probability inequalities of \cite{den2022} which allow similar bounds to those of martingale difference process, for mixing processes. Finally, \ref{ass2} specifies exponential probability tails for the regressors and errors. This is a reasonably strong assumption that is common in the high dimensional regression literature and is needed again for probability inequalities. It can be replaced by a polynomial tail assumption as discussed extensively  in \cite{den2022}. It would require a restriction on the value of $\kappa_1$ in Proposition 1. Exponential probability tails allow that $N$ is an exponential function of $T$. Since we do not restrict $\kappa_1$ we essentially allow that $N$ rises exponentially with $T$ and we therefore need this assumption.
\end{remark}

\end{document}